\documentclass[final,5p,times,twocolumn]{elsarticle}

\usepackage{amssymb}
\usepackage{amsthm}
\usepackage{amsmath}

\journal{Physica B: Condensed Matter}

\begin{document}

\begin{frontmatter}



\title{Unusual critical currents in quasi-one-dimensional superconducting aluminum two-width structures in a magnetic field}


\author{V.~I.~Kuznetsov \corref{cor1}}
\ead{kuznetcvova@mail.ru}
\cortext[cor1]{corresponding author}
\author{O.~V.~Trofimov}
\address{Institute of Microelectronics Technology and High
Purity Materials, Russian Academy of Sciences, Chernogolovka,
Moscow Region 142432, Russia}

\begin{abstract}
{We measured critical switching and retrapping currents as
functions of temperature in the zero field and as functions of a
magnetic field perpendicular to the substrate surface at a
temperature close to the critical temperature in thin-film long
quasi-one-dimensional superconducting aluminum two-width
structures consisting of narrow and wide wires with different
critical temperatures. It is found that the switching current as a
function of the field is nonlocal (dependent on electron transport
in the area containing the junction line between the narrow and
wide wires). The unusual behavior of the switching current lies in
the fact that it is poorly described by the classical
Ginzburg-Landau theory. In low and intermediate fields, the
switching current is determined mainly by the densities of
critical currents in the narrow and wide wires, respectively, near
the wire junction line. In high fields, the switching current is
due to equilibrium fluctuations of the order parameter.}
\end{abstract}




\end{frontmatter}



\section{Introduction}

Miniaturization of electron superconducting and hybrid structures
and microdevices leads to the manifestation of new nonequilibrium
and nonlocal effects \cite{art1}, \cite{art2},
\cite{kuznphysicacnegativ20}. Nonlocal effects can be useful for
creating microdevices. For example, the nonlocal effect leading to
spin entanglement of spatially separated electrons can be used to
create quantum computing \cite{art1}. Nonequilibrium and nonlocal
effects can be undesirable, degrading the performance of
microdevices. In particular, nonequilibrium quasiparticle
excitations arising from the depairing of superconducting pairs
\cite{art4} create problems for the operation of superconducting
radiation detectors and qubits \cite{art5}.

The study of electron transport and nonequilibrium and nonlocal
effects in quasi-one-dimensional superconducting wires and hybrid
structures is relevant, since such wires are often key elements of
microdevices. In particular, a superconducting aluminum Dayem
nanobridge was used for nanosecond thermometry \cite{art6}.
Superconducting nanowires are key elements of the asymmetric
nanowire superconducting quantum interference device (SQUID)
\cite{art7}.

Electron transport in quasi-one-dimensional superconducting wires
is strongly dependent on fluctuations. At temperatures slightly
below the critical temperature $T{c}$ and currents close to the
critical current and low currents, thermal nonequilibrium
fluctuations give rise to so-called "phase slip centers (PSCs)"
and "thermally activated phase slips (TAPSs)," respectively
\cite{art8}, \cite{art9}. In PSCs, the superconducting order
parameter periodically vanishes, with the phase difference at the
PSC varying by 2$\pi$. In PSCs, dissipation and superconductivity
coexist. At very low temperatures in a wire with an effective
diameter smaller than 10 nm, quantum fluctuations of the order
parameter cause quantum phase slips (QPSs) \cite{art9},
\cite{art10}. QPSs switch the wire from the superconducting state
to the resistive or insulating state. At temperatures above
$T{c}$, equilibrium fluctuations lead to a decrease in resistance
\cite{art11}.

Coherent QPSs can be used as key elements for phase-slip flux
qubits \cite{art12}, \cite{art13}, \cite{art14}. Superconducting
titanium nanowires containing quantum phase slips (QPSs) are used
to replace tunnel junctions in a Cooper pair transistor
\cite{art15}.

Despite the numerous works related to quasi-one-dimensional
superconducting wires (see review papers \cite{art9},
\cite{art10}), open questions remain. Such effects as the
"resistive transition anomaly" \cite{art16}, \cite{art17},
\cite{art18}, \cite{art19}, \cite{art20}, \cite{art21},
\cite{art22}, \cite{art23}, the negative magnetoresistance in low
magnetic fields \cite{art24}, \cite{art25}, \cite{art26}, and the
"antiproximity" effect \cite{art27}, \cite{art28} have no
generally accepted explanation. The "resistive transition anomaly"
manifests itself in the $R(T)$ dependence during the transition
from the normal state to the superconducting state as a surge in
resistance above the normal resistance. The "antiproximity" effect
and related effects consist of the fact that previously suppressed
superconductivity of nanowires is restored when the order
parameter in large superconducting contact pads decreases upon the
application of a magnetic field \cite{art27}, \cite{art28},
\cite{art29}.

The critical current of a superconducting quasi-one-dimensional
wire has its own characteristics \cite{art30}. The critical
current at which the dc voltage appears with increasing direct
current is called the switching current. Retrapping current is the
critical current at which the dc voltage disappears when the
direct current decreases.

In our work, we experimentally study the critical current of a
thin-film long (with a length several times exceeding the length
of the quasiparticle imbalance $\lambda_{Q}(T,B)$, depending on
temperature $T$ and magnetic field $B$) superconducting
quasi-one-dimensional aluminum structure consisting of narrow and
wide wires with different critical temperatures. Aluminum was
chosen as the material for the structure due to its long
superconducting coherence length, easily achievable critical
temperature by pumping $^{4}$He vapor, and the fact that
superconducting quasi-one-dimensional aluminum wire is often a
component of hybrid structures and microdevices \cite{art31},
\cite{art32}, \cite{art33}.

The authors of \cite{art31} found that different "global and
bimodal superconducting states" arise in a superconducting
aluminum nanowire whose ends are attached to large normal contact
pads when this wire is driven into a nonequilibrium state by an
external current. In \cite{art32}, two-electron tunneling local
Andreev, non-local Andreev and elastic processes were measured
using a Cooper pair splitter, one of the key elements of which is
a superconducting aluminum wire in contact with two copper islands
through a tunnel barrier \cite{art33}.

New effects in electron transport were observed in
\cite{kuznjetpletmar16}, \cite{kuznphysicaccrit22},
\cite{kuznprb08}, \cite{kuznphysica13}, \cite{kuznphysicacshift22}
thin-film long quasi-one-dimensional superconducting aluminum
structures of equal thicknesses and consisting of narrow and wide
wires, at temperatures slightly below the critical temperature
$T_{c}$ in the zero and non-zero magnetic fields perpendicular to
the substrate surface.

The trigger for the above studies was the work
\cite{dubjetplet03},  in which a nonzero rectified time-averaged
direct voltage was measured in a superconducting circularly
asymmetric aluminum ring and structures consisting of such rings
permeated with a magnetic flux and biased with a sinusoidal
current (with a zero dc component) with an amplitude close to the
critical value. This voltage, as a function of the magnetic field,
oscillates with periods corresponding to the superconducting
quantum of the magnetic flux. The voltage appears due to a
mysterious "shift of the maxima of critical currents of different
polarity relative to the zero flux in opposite directions along
the flux axis in the asymmetric ring." It was found in
\cite{kuznphysicacshift22} that this shift is nonzero only in the
case of different critical temperatures of the narrow and wide
wires that make up the ring. The superconducting aluminum ring,
consisting of narrow and wide wires, and a series of such rings is
a very effective ac voltage rectifier \cite{dubjetplet03} and can
work as an asymmetric micro-SQUID \cite{art7} and a highly
sensitive detector of electromagnetic nonequilibrium noise.

It was found in \cite{kuznjetpletmar16} that the $I$-$V$ curves of
aluminum structures, consisting of narrow and wide wires, measured
in a perpendicular magnetic field, contain areas of almost
constant voltage, which are subharmonics of the superconducting
gap. These subharmonics are due to the multiple Andreev reflection
in the overheated nonequilibrium section of the structure.

Negative local and nonlocal voltages (resistances), dependent on
temperature and a perpendicular magnetic field, were measured in
an aluminum structure composed of narrow and wide wires. This
effect occurs in a certain temperature range when the narrow and
wide wires are normal and superconducting, respectively
\cite{kuznphysicacnegativ20}.

It was assumed that the electronic transport in these aluminum
structures is due to the fact that the structures under study
consist of narrow and wide wires with different critical
temperatures.

Indeed, the authors of  \cite{kuznphysicaccrit22} experimentally
demonstrated that the critical temperature of a narrow wire is
lower than that one of a wide wire. Furthermore, the critical
switching and retrapping currents were measured as functions of
temperature in two aluminum structures at temperatures just below
the critical temperature $T_{c}$ in the zero magnetic field. The
dc voltage was recorded on the central short section of these
structures. It was found that at a given temperature, the
switching current density of a narrow wire is less than that one
of a wide wire. The experimental switching current is nonlocal
(dependent on the physical parameters of both wide and narrow
wires) and is fitted to two functions \cite{kuznphysicaccrit22}.

The critical current as a function of magnetic field in such
aluminum structures consisting of narrow and wide wires has not
been studied either theoretically or experimentally. We expect the
switching current as a function of magnetic field to be a nonlocal
quantity, determined not only by the electron transport properties
of the central section of the structure but also by the electron
transport properties of other sections of the wide and narrow
wires. At a given temperature in low fields, the switching current
density in the narrow wire of the structure is expected to be less
than that one in the wide wire. However, in a certain intermediate
range of magnetic fields, an interesting case may be realized
where the switching current density in the narrow wire is greater
than that one in the wide wire. This may occur because the
superconducting order parameter is suppressed by the magnetic
field more strongly, the wider the wire \cite{tinkham}.

In this work, we measured the critical currents as functions of
temperature $T$ close to $T_{c}$ in the zero magnetic field in
three thin-film long quasi-one-dimensional superconducting
aluminum structures of other geometries. For this purpose, we
recorded $I$-$V$ curves with applied increasing and decreasing
direct currents at different temperatures $T$ in the zero magnetic
field. The dc voltage was recorded from the central short section
of the structures. $I$-$V$ curves exhibit hysteresis.

We do not show here the $I$-$V$ curves measured at different
temperatures in the zero field. We suggest taking a look at the
branches of the $I$-$V$ curves from \cite{kuznjetpletmar16}, which
are measured with decreasing applied current in aluminum
structures very similar to the structures studied here.

The critical switching current abruptly transfers the structure
from a superconducting state to a resistive state close to the
normal state. In our case, the critical switching current
coincides with the depairing Ginzburg-Landau critical current
$I_{c}(T)$ in a single wire of constant width. The transition from
a resistive state to a superconducting state with decreasing
applied current is current-stretched and contains non-monotonic,
almost linear sections and sections of almost constant voltage.
Non-monotonic sections of the $I$-$V$ curves, manifesting as
maxima and minima, are due to the competition between overheating
and cooling of the nonequilibrium region located at the phase slip
center or self-forming SNS contact. Nearly linear sections of the
$I$-$V$ curves and excess superconducting current indicate the
presence of a phase slip center. Almost constant voltage sections
occur due to multiple Andreev reflections in the phase slip
centers or SNS junctions formed in the structures
\cite{kuznjetpletmar16}.

In addition, at a given temperature $T_{g}$ close to $T_{c}$, we
measured $I$-$V$ curves in a magnetic field $B$ perpendicular to
the substrate surface in two previously studied structures
\cite{kuznphysicaccrit22} and in three other structures using
different measurement circuits. $I$-$V$ curves recorded in the
magnetic field also have hysteresis depending on the increase and
decrease of the direct current. We studied the magnetic field
dependences of the switching $I_{c}(T_{g},B)$ and retrapping
$I_{r}(T_{g},B)$ currents obtained from the $I$-$V$ curves. At low
temperatures and low fields, the switching current is greater than
the retrapping current. At temperatures and fields close to the
critical values, both currents coincide.

We propose to look at the branches of the $I$-$V$ curves from
\cite{kuznjetpletmar16} recorded with decreasing applied current
in aluminum structures placed in a perpendicular magnetic field at
a given temperature.

\section{Methods}

We used a PMMA/P(MMA-8.5MAA) two-layer resist lift-off  electron
beam lithography to fabricate the aluminum structures. The PMMA
(polymethyl methacrylate) is used as a less sensitive top layer.
The P(MMA-8.5MAA) is a copolymer of methyl methacrylate (MMA) and
methacrylic acid (MAA) with an 8.5 \% methacrylic acid content,
which is used as a more sensitive bottom layer. The "NanoMaker"
software was used to correct the proximity effect in electron beam
lithography.

Deposition of a thin aluminum film with the thickness $d$ occurred
via thermal evaporation onto a silicon substrate at room
temperature at a rate of 1 nm/s. The vacuum chamber pressure was
$5\times 10^{-7}$ Torr. The substrate was located at a distance of
10 cm from the source of evaporated aluminum. The purity of the
aluminum used for film deposition was 99.999 \%. The films were
polycrystalline with a crystallite size of 25 nm. The
root-mean-square roughness of the films was 2 nm. The silicon
substrates were $5\times 5$ mm$^{2}$ square chips with
pre-deposited gold contacts. The chips were cut from a (100)
oriented silicon wafer, as such wafers are easily separated into
rectangular fragments.

The $I$-$V$ curves were measured taking all possible precautions
to minimize the influence of electromagnetic noise and
interference (see \cite{kuznphysicaccrit22} for details).
Measurement errors for current and voltage did not exceed 10 \% of
the measured values. The $I$-$V$ curves were recorded three times
at a given temperature and magnetic field. Average values of the
critical switching and retrapping currents were taken.

The temperature was regulated by pumping out $^{4}$He vapor.
Temperature was determined by the saturated vapor pressure and a
calibrated carbon resistor thermometer with a value of 1.121
k$\Omega$ at room temperature. The thermometer was powered by a
direct current of 0.1 $\mu$A from a battery of galvanic cells.
Temperature measurement errors did not exceed ten percent of the
measured value. The thermometer and the chip containing the
structure were mounted on a copper sample holder located at the
lower end of an insert placed inside a glass cryostat containing
liquid $^4$He.

To create a uniform magnetic field perpendicular to the substrate
surface, small Helmholtz coils were used. The coil windings were
made of superconducting NbTi wire covered with a copper sheath
with an outer diameter of 0.33 mm. The coils were attached to the
lower end of the outer insert in a glass cryostat with liquid
$^{4}$He. The magnetic field magnitude $B$ was determined by the
linear dependence $B=KI_{m}$, where $K=98$ G/A is the
proportionality constant and $I_{m}$ is the external direct
current flowing through the Helmholtz coils. The magnitude of $K$
was found from magnetic field measurements with a calibrated Hall
sensor and corrected based on measurements of quantum periodic
voltage oscillations in a superconducting aluminum ring with a
diameter of 4 $\mu$m, biased by a small direct current near the
critical temperature and threaded by a magnetic flux. Errors in
determining the magnetic field using the linear dependence did not
exceed five percent of the measured value. The linear dependence
$B=KI_{m}$ was unambiguous without hysteresis when the current
$I_{m}$ slowly ramped from -1.45 A to 1.45 A and back. In this
current range, the linear dependence showed no field jumps or
signs of magnetic flux trapping in the superconducting winding. A
permalloy shield was used to attenuate the external magnetic
field.

\section{Results and discussion}

\subsection{Geometrical and physical parameters of the five structures}

\begin{figure}
\begin{center}
\includegraphics[width=1\linewidth]{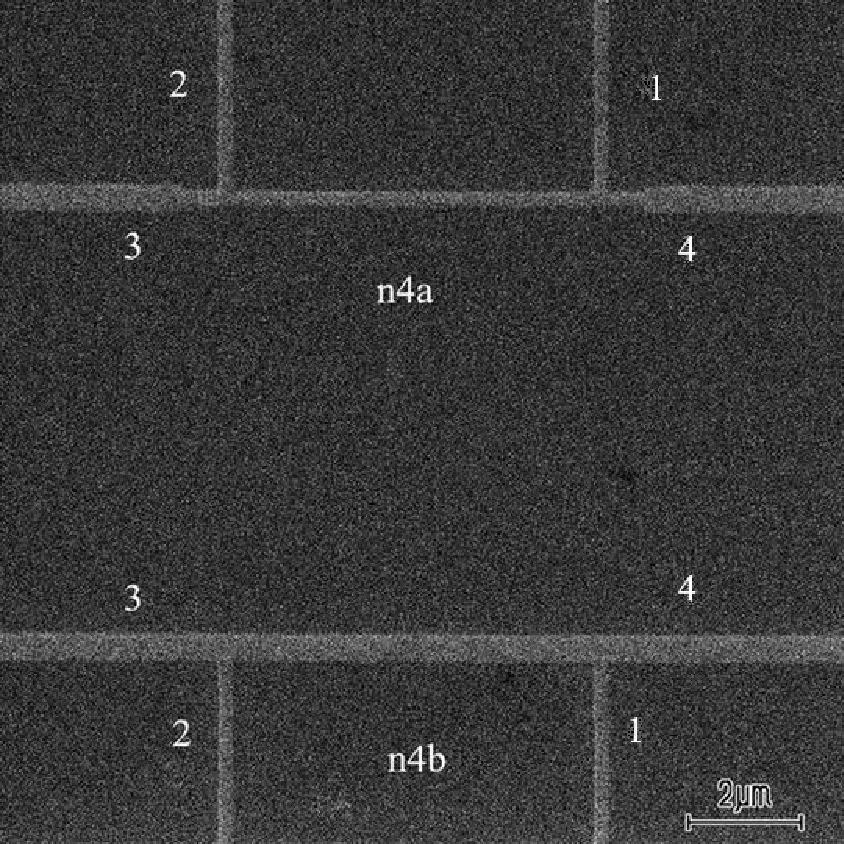}
\caption{\label{f1} (Color online) SEM image of the central parts
of the "n4a" and "n4b" structures. The numbers 1, 2, 3, and 4
refer to the contact numbers used in the measurement circuits
(Table 2). The horizontal scale bar is 2 $\mu$m.}
\end{center}
\end{figure}

Fig. \ref{f1} shows scanning electron microscope image of the
central parts of the "wide" and "narrow" structures previously
used in \cite{kuznphysicaccrit22} and hereinafter referred to as
"n4b" and "n4a", respectively. A distinctive feature of the "n4b"
and "n4a" structures is that they are fabricated on the same chip
and have the same thickness. In addition, the widths of the narrow
and wide wires that make up both structures are approximately the
same, $w_{n}$ = 0.27 and $w_{w}$ = 0.48 $\mu$m, respectively. The
narrowing is in the center of the "n4a" structure. The "n4b" and
"n4a" structures contain narrow and wide wires with two different
widths, so we will call them "two-width structures".

Table \ref{t1} presents the geometrical and physical parameters of
the "n4b" and "n4a" structures (Fig. \ref{f1}) and other "n1"
(inset, Fig. \ref{f2}), "sl5" (inset, Fig. \ref{f3}(a)), "sl6"
(inset, Fig. \ref{f3}(b)) structures. All five structures consist
of long (with a total length that can reach 70 $\mu$m)
quasi-one-dimensional superconducting narrow and wide wires. The
large length of the wires allows reducing the influence of wide
($>5$ $\mu$m) superconducting current and potential contacts on
electron transport in the central part of the structure. The
superconducting wire is quasi-one-dimensional because twice the
temperature-dependent superconducting coherence length $2\xi(T)$
is greater than the wire width $w$. In addition, the coherence
length $2\xi(T)$ is significantly greater then the film thickness
$d$. For all structures in the temperature range under study, the
quasi-one-dimensionality condition is satisfied.

The "n1" structure (inset, Fig. \ref{f2}) consists of
quasi-one-dimensional superconducting current and potential wires
of the same width. A feature of the central section of the "sl5"
structure (inset, Fig. \ref{f3}(a)) is that the length of the
narrow short wire from which the dc voltage is recorded is
approximately equal to the temperature-dependent superconducting
coherence length $\xi(T)$ for the studied temperature range near
$T_{c}$. Moreover, the junction line of the wide and narrow wires
is located practically next to the nearest potential wire. The
"sl5" structure contains narrow and wide wires with two different
widths, so we will classify it as "two-width structure".

In order to find the expected long-range (magnetic
field-dependent) influence of electron transport near the junction
line of the wide and narrow current wires on the electron
transport in the central section of the narrow wire of length
approximately equal to $6\xi(T)$, we fabricated the "sl6"
structure (inset, Fig. \ref{f3}(b)). In the "sl6" structure, the
junction lines of the wide and narrow current wires are located at
distances approximately equal to $16\xi(T)$ and $5\xi(T)$ from the
nearest potential wires in the cases ($I$-3-4, $V$-2-1) and
($I$-2-1, $V$-3-4), respectively. The narrow wires of the
structure have the same width. The widths of the wide wires of the
structure are specially made different (Table \ref{t1}) to compare
the (magnetic field-dependent) influences of the wide wires of the
different widths on the electron transport in the central section
of the narrow wire. The "sl6" structure consists of wires with
three different widths, so we will classify it as "three-width
structure".

The width of the central section (narrowing) of the structure is
denoted, where the dc voltage is measured, as $w_{a}$ and the
width of other sections of the structure as $w_{b}$. Furthermore,
in this work, the widths of the narrow and wide wires of the
structure are denoted as $w_{n}$ and $w_{w}$, respectively.

\begin{table*}
\caption{\label{t1} Geometrical and physical parameters for five
structures: $l_{V}$ is the length of the central section
(narrowing) of the structure, on which the dc voltage $V$ is
recorded, $w_{a}$ is the width of the central section of the
structure, $w_{b}$ is the width of other sections of the
structure, $d$ is the film thickness, $R_{n}$ is the resistance of
the central section of the structure in the N state, $\rho_{n}$ is
the resistivity in the N state, $l_{el}$ is the electron mean free
path, $\xi(0)$ is the superconducting coherence length and
$\lambda(0)$ is the magnetic field penetration depth at $T=0$\,K,
$T_{c}(0.5)$ is the critical temperature determined at the middle
of the resistive N-S transition, $j_{GL}(0)$ and $j_{KL}(0)$ are
the critical current densities at $T=0$\,K in $B=0$\,G, calculated
within the framework of the Ginzburg-Landau (GL) and
Kupriyanov-Lukichev (KL) theories, respectively.}
\centering %
\begin{tabular}{ccccccccccccc}
\hline
 & $l_{V}$, & $w_{a}$, & $w_{b}$, & $d$, & $R_{n}$, & $\rho_{n}$, & $l_{el}$, & $\xi(0)$, & $\lambda(0)$, & $T_{c}(0.5)$, & $j_{GL}(0)$, & $j_{KL}(0)$,\\
 & $\mu$m & $\mu$m & $\mu$m & nm &  $\Omega$ &  $10^{-8}\Omega$\,m &  nm &  $\mu$m &  $\mu$m & K & $10^{10}$\,A/m$^{2}$  & $10^{10}$\,A/m$^{2}$\\
\hline
n4b & 6.69 & 0.48 & 0.27 & 19 & 37.9 & 5.2 & 9.9 & 0.107 & 0.125 & 1.485 & 6.04 & 7.60\\
n4a & 6.69 & 0.27 & 0.48 & 19 & 69.0 & 5.3 & 9.6 & 0.105 & 0.127 & 1.455 & 5.96 & 7.20\\
n1 & 7.99 & 0.46 & 0.46 & 32 & 16.9 & 3.1 & 16.4 & 0.138 & 0.096 & 1.407 & 8.05 & 8.97\\
sl5 & 1.01 & 0.23 & 0.40 & 30 & 5.4 & 3.6 & 14.2 & 0.128 & 0.103 & 1.355 & 7.50 & 7.89\\
sl6 & 6.03 & 0.20 & 0.39 (0.60) & 30 & 42.4 &  4.1 & 12.4 & 0.120 & 0.110 & 1.463 & 7.01 & 8.29\\
\hline
\end{tabular}
\end{table*}

We found the electron mean free path $l_{el}$ from the refined
theoretical expression $\rho l_{el}=5.1 \times 10^{-16}$ $\Omega$
m$^{2}$ \cite{gershenson}. For the structures under study,  the
dirty limit is valid, since $l_{el}$ is much less than the
superconducting coherence length of pure aluminum $\xi_{0}=1.6$
$\mu$m.

For the dirty limit at $T$ slightly below $T_{c}$, the
Ginzburg-Landau coherence length as a function of temperature is
given by the expression $\xi(T)=\xi(0)(1-T/T_{c})^{-1/2}$, where
$\xi(0)=0.85(l_{el}\xi_{0})^{1/2}$ is the coherence length at
$T=0$\,K \cite{schmidt}. The magnetic field penetration depth as a
function of temperature is determined by the expression
$\lambda_{GL}(T)=\lambda(0)(1-T/T_{c})^{-1/2}$, where
$\lambda(0)=0.615\lambda_{L}(\xi_{0}/l_{el})^{1/2}$ is the field
penetration depth at $T=0$\,K, $\lambda_{L}=16$ nm is the London
penetration depth for aluminum \cite{schmidt}.

The Ginzburg-Landau depairing critical current density of the as a
function of temperature is determined by the expression
$j_{GL}(T)=j_{GL}(0)(1-T/T_{c})^{3/2}$ \cite{schmidt}, where
$j_{GL}(0)=c\Phi_{0}/(12\sqrt{3}\pi^{2}\lambda(0)^{2}\xi(0))$ is
the critical current density at $T=0$\,K, $\Phi_{0}$ is the
superconducting quantum of magnetic flux. In addition, at $T$
below $T_{c}$, to calculate the critical current density as a
function of temperature in a dirty quasi-one-dimensional
superconducting wire, we used the expression
$j_{c}(T)=(j_{KL}(0)/4)(1-(T/T_{c})^{2})(1-(T/T_{c})^{4})^{1/2}$,
obtained in \cite{romijn} within the Kupriyanov-Lukichev theory
\cite{kupriyanov}, where
$j_{KL}(0)=((8\pi^{2}\sqrt{2\pi})/(21\zeta(3)e))\sqrt{(kT_{c})^{3}/(\hbar
v_{F}\rho_{n} \rho_{n}l_{el})}=(9.493\times10^{6}(T_{c}^{3}/
\rho_{n})^{1/2})$ A/m$^{2}$ is the critical current density at
$T=0$\,K. In the calculation, we took $T_{c}=T_{c}(0.5)$, where
$T_{c}(0.5)$ is the critical temperature corresponding to the
middle of the resistive N-S transition of the wire biased by the
low dc $I_{dc}=0.1$ $\mu$A.

\subsection{Switching and retrapping currents vs. $T$ measured in "n1", "sl5", "sl6" structures near the critical temperature at $B=0$\,G. Theoretical and fitted parameters used to fit the experimental switching and retrapping currents vs. $T$ for five structures "n4b", "n4a", "n1", "sl5", "sl6".}

\begin{figure}
\begin{center}
\includegraphics[width=1\linewidth]{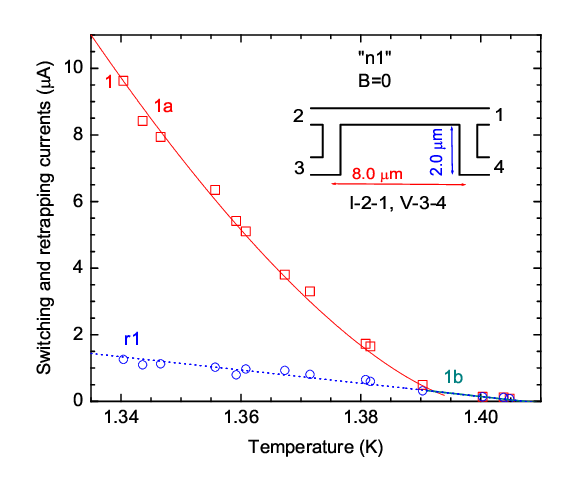}
\caption{\label{f2} (Color online) Experimental data for the "n1"
structure: 1 (squares) and r1 (circles) are the switching and
retrapping currents as functions of $T$ at $B=0$ G, respectively.
Solid lines 1a and 1b are the fits of the measured switching
current (squares) to the functions $I_{GL}(T)$ (\ref{e2}) at low
$T$ and $I_{r}(T)$ (\ref{e4}) at high $T$ , respectively. The
dotted line is the fit of the measured retrapping current
(circles) to the function $I_{r}(T)$ (\ref{e4}). The inset: a
sketch of the structure (not to scale) with the measurement
circuit.}
\end{center}
\end{figure}

\begin{figure}
\begin{center}
\includegraphics[width=1\linewidth]{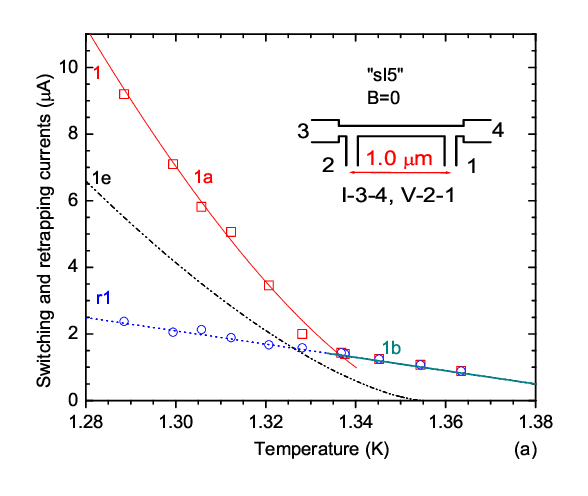}
\includegraphics[width=1\linewidth]{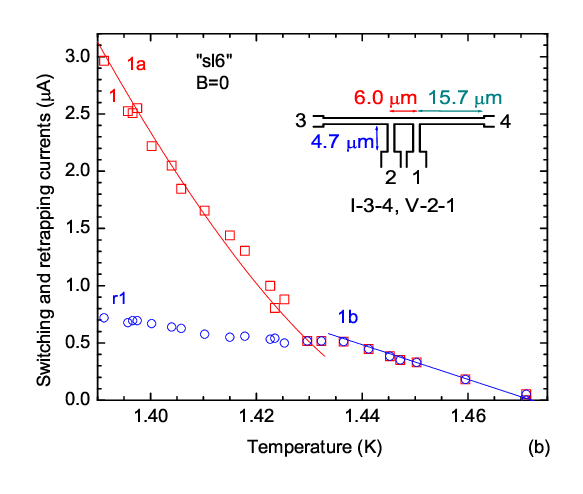}
\caption{\label{f3} (Color online) (a)-(b). Experimental data for
the "sl5" \, ($I$-3-4, $V$-2-1) in (a) and "sl6" \, ($I$-3-4,
$V$-2-1) in (b) structures: 1 (squares) and r1 (circles) are the
switching and retrapping currents as functions of $T$ at $B=0$ G,
respectively. (a)-(b) Solid lines 1a and 1b are the fits of the
measured switching current (squares) to the functions $I_{GL}(T)$
(\ref{e2}) at low $T$ and $I_{r}(T)$ (\ref{e4}) at high $T$ ,
respectively. (a) Dash-dotted line 1e is the expected switching
current as a function of $T$ calculated using the function
$I_{GL}(T)$ (\ref{e1}). (a) The dotted line is the fit of the
measured retrapping current (circles) to the $I_{r}(T)$ function
(\ref{e4}). (a) - (b) The insets: sketches of the structures (not
to scale) with the measurement circuits.}
\end{center}
\end{figure}

Figs. \ref{f2}, \ref{f3}(a), and \ref{f3}(b) show the critical
switching (1 - squares) and retrapping (r1 - circles) currents vs.
$T$, measured in the "n1", "sl5", and "sl6" structures near the
critical temperature in the zero field. Solid lines 1a and 1b are
the fits of the switching current at lower and higher
temperatures, respectively. Dotted lines are the fits of the
retrapping current. The insets in the figures contain sketches of
the central sections of the structures (not to scale) with
different measurement circuits. Dash-dotted line 1e (Fig.
\ref{f3}(a)) demonstrates the switching current vs. $T$ calculated
for the individual narrow wire of the "sl5" structure.

At lower temperatures, the switching current is higher than the
retrapping current. At higher temperatures, the switching current
coincides with the retrapping current, which depends linearly on
temperature (Figs. \ref{f2}, and \ref{f3}). It can be seen that
for the more uniform constant-width "n1" structure, line 1b is
very short. This means that the experimental switching current is
fitted to the single line 1a, in almost the entire temperature
range under study (Fig. \ref{f2}).

We found that for the "n1" and "sl6" structures, the measured
critical switching current at a given temperature coincides with
the temperature-dependent critical depairing Ginsburg-Landau
current $I_{GL}(T)$.

Table \ref{t2} contains theoretical and fitted parameters used to
fit the experimental switching and retrapping currents vs. $T$
near the critical temperature at $B=0$ G for five structures
"n4b", "n4a", "n1", "sl5", "sl6".

\begin{table*}
\caption{\label{t2} Theoretical and fitted parameters used in the
equations $I_{GL}(T)$ (\ref{e2}), $I_{KL}(T)$ (\ref{e3}) and
$I_{r}(T)$ (\ref{e4}) to fit the measured switching and retrapping
currents vs. $T$ for five structures: circuit is a measurement
circuit, $T_{1}$ is the temperature range for switching current,
$I_{GL}(0)$ is the GL critical current at $T=0$\,K,
$I_{GL}^{f}(0)$ is the fitted GL critical current at $T=0$ K,
$T_{GL}^{cf}$  is the fitted critical temperature, $I_{KL}(0)$  is
the KL critical current at $T=0$\,K, $I_{KL}^{f}(0)$ is the fitted
KL critical current at $T=0$\,K, $T_{KL}^{cf}$ is the fitted
critical temperature, $T_{2}$ is the temperature range for
retrapping current, $I_{rf}(0)$ is the fitted retrapping current
at $T=0$ K, $T_{r}^{cf}$ is the fitted critical temperature.
$T_{c}(0.5)$ is the critical temperature determined in the middle
of the resistive N-S transition.}
\centering %
\begin{tabular}{ccccccccccccc}
\hline
 & circuit & $T_{1}$, & $I_{GL}(0)$, & $I_{GL}^{f}(0)$, & $T_{GL}^{cf}$, & $I_{KL}(0)$, & $I_{KL}^{f}(0)$, & $T_{KL}^{cf}$, & $T_{2}$, & $I_{rf}(0)$, & $T_{r}^{cf}$, & $T_{c}(0.5)$, \\
 & & K & $\mu$A & $\mu$A & K & $\mu$A & $\mu$A & K & K &  $\mu$A & K  & K\\
\hline
n4b & $I$-3-4, & 1.250- & 550 & 600 & 1.472 & 690 & 732 & 1.466 & 1.250- & 22 & 1.510 & 1.485\\
& $V$-2-1 & 1.466 &  &  &  &  &  &  & 1.445 &  &  &\\
n4b & $I$-2-1, & 1.250- & 306 & 320 & 1.437 & 371 & 408 & 1.427 & 1.250-  & 17 & 1.525 & 1.487\\
& $V$-3-4 & 1.400 &  &  &  &  &  &  & 1.475 &  &  &\\
n4a & $I$-3-4, & 1.250- & 306 & 320 & 1.437 & 371 & 408 & 1.427 & 1.250- & 14 & 1.457 & 1.455\\
& $V$-2-1 & 1.410 &  &  &  &  &  &  & 1.450 &  &  &\\
n1 & $I$-2-1, & 1.340- & 1190 & 1150 & 1.398 & 1327 & 1200 & 1.398 & 1.340- & 28 & 1.407 & 1.407\\
& $V$-3-4 & 1.390 &  &  &  &  &  &  & 1.407 &  &  &\\
sl5 & $I$-3-4, & 1.289- & 893 & 860 & 1.355 & 940 & 900 & 1.355 & 1.289- & 28 & 1.405 & 1.355\\
& $V$-2-1 & 1.337 &  &  &  &  &  &  & 1.363 &  &  &\\
sl6 & $I$-3-4, & 1.321- & 410 & 400 & 1.447 & 485 & 428 & 1.447 & 1.437- & 22 & 1.472 & 1.463\\
& $V$-2-1 & 1.430 &  &  &  &  &  &  & 1.472 &  &  &\\
\hline
\end{tabular}
\end{table*}

Near $T_{c}$, the critical current vs. $T$ is given by the
equation
\begin{equation}
\label{e1} I_{GL}(T)=I_{GL}(0)(1-T/T_{c})^{3/2},
\end{equation}
where $I_{GL}(0)=j_{GL}(0)dw_{a}$ is the critical current at
$T=0$\,K, $T_{c}=T_{c}(0.5)$. In the temperature range $T_{1}$, we
fitted the experimental switching current vs. $T$ within the GL
model to the equation
\begin{equation}
\label{e2} I_{GL}(T)=I_{GL}^{f}(0)(1-T/T_{GL}^{cf})^{3/2},
\end{equation}

where $I_{GL}^{f}(0)$ is the fitted GL critical current at
$T=0$\,K and $T_{GL}^{cf}$ is the fitted critical temperature
(Table \ref{t2}). We found that $I_{GL}^{f}(0)$ differs from
$I_{GL}(0)$ by 3-4 \% and $T_{GL}^{cf}$ is usually smaller than
$T_{c}(0.5)$ (Table \ref{t2}).

In addition, the measured switching current vs. $T$ is fitted
within the framework of the Kupriyanov-Lukichev theory to the
equation
\begin{equation}
\label{e3}
I_{KL}(T)=(I_{KL}^{f}(0)/4)(1-(T/T_{KL}^{cf})^{2})(1-(T/T_{KL}^{cf})^{4})^{1/2},
\end{equation}
where $I_{KL}^{f}(0)$ is the fitted KL critical current at
$T=0$\,K, $T_{KL}^{cf}$ is the fitted critical temperature (Table
\ref{t2}). We found that $I_{KL}^{f}(0)$ differs from the
theoretical KL critical current at $T=0$\,K, equal to
$I_{KL}(0)=j_{KL}(0)dw_{a}$ by 4-13 \% and $T_{KL}^{cf}$ is
usually less than $T_{c}(0.5)$ (Table \ref{t2}).

Note that for the "sl5" structure (Fig. \ref{f3}(a)) with a very
short narrowing, the measured switching current values lie almost
twice as high as the dash-dotted line 1e in the range of
1.289-1.337 K. Line 1e shows the expected switching current vs.
$T$ calculated for the individual narrow wire of the structure
with the width $w_{a}=w_{n}=0.23$ $\mu$m using the function
$I_{GL}(T)$ (\ref{e1}). The current flows through the "sl5"
structure without noticing the narrowing, therefore $I_{GL}(0)$
and $I_{KL}(0)$ are calculated for the individual wide wire of the
structure with the width $w_{b}=w_{w}=0.40$ $\mu$m. In this case,
the switching current density in the narrowing exceeds the GL
critical current density $j_{GL}(0)$ almost twice. The fitted
curves $I_{GL}(T)$ (\ref{e2}) and $I_{KL}(T)$ (\ref{e3})
practically coincide, therefore only the curves $I_{GL}(T)$ (solid
lines 1a) are placed in Figs. \ref{f2}, \ref{f3}(a), and
\ref{f3}(b).

At higher temperatures, the solid line 1b and the dotted line,
approximating the temperature dependences of the measured
switching and retrapping currents, respectively, coincide (Figs.
\ref{f2}, and \ref{f3}). Therefore, in the higher temperature
range, the experimental switching current is fitted by the same
function as the retrapping current. In the temperature range
$T_{2}$, the measured retrapping current vs. temperature fitted to
the equation
\begin{equation}
\label{e4} I_{r}(T)=I_{rf}(0)(1-T/T_{r}^{cf})
\end{equation}
(dotted lines in Figs. \ref{f2}, and \ref{f3}(a)), where
$I_{rf}(0)$ is the fitted retrapping current at $T=0$ K,
$T_{r}^{cf}$ is the fitted critical temperature, which is usually
higher than the critical temperature $T_{c}(0.5)$ (Table
\ref{t2}).

At lower temperatures, we believe that the linear temperature
dependence of the retrapping current is due to Joule (or
quasiparticle) overheating of the structure. At higher
temperatures, where the switching and retrapping currents
coincide, we believe that the "n4b" and "n4a" structures
\cite{kuznphysicaccrit22} and the "sl5" and "sl6" structures,
measured according to certain electrical circuits, are Josephson
SNS structures. This is due to the fact that the critical
temperature $T_{cn}$ of the individual (without taking into
account the influence of the proximity effect) narrow wire of the
structure is lower than the critical temperature $T_{cw}$ of the
individual (without taking into account the influence of the
proximity effect) wide wire of the structure
\cite{kuznphysicaccrit22}. We assume that in the case of higher
temperatures ($T_{cn}<T<T_{cw}$), the linear temperature
dependence is determined by the linear dependence of the Josephson
critical current $I_{J}(T)$ \cite{likharev}.

In \cite{kuznphysicaccrit22} it was found that the critical
switching and retrapping currents coincide in the "n4b" structure
measured according to the $I$-2-1, $V$-3-4 scheme at
$T=1.404-1.475$ K. In this temperature range, the critical current
is fitted to the $I_{r}(T)$ function (\ref{e4}) with the fitted
retrapping current $I_{rf}(0)=17$ $\mu$A and the fitted critical
temperature $T_{r}^{cf}=1.525$ K (Table \ref{t2}). We consider
that $T_{cn}$ is close to the temperature of 1.404 K, below which
the switching current becomes higher then the retrapping current.
We believe that $T_{r}^{cf}$ is close to $T_{cw}$ for the "n4b"
($I$-2-1, $V$-3-4) and "sl5" (Fig. \ref{f3}(a)), "sl6" (Fig.
\ref{f3}(b)) structures.

In \cite{kuznphysicaccrit22}, it was measured that the switching
and retrapping currents are the same in the "n4a" structure
($I$-3-4, $V$-2-1) at $T=1.415-1.444$ K. At these temperatures,
the critical current is fitted to the function $I_{r}(T)$
(\ref{e4}) with $I_{rf}(0)=14$ $\mu$A and $T_{r}^{cf}=1.457$ K
(Table \ref{t2}). We consider that the temperatures of 1.415 and
1.457 K are close to $T_{cn}$ and $T_{cw}$, respectively.
Therefore, at $T=1.415-1.457$ K, the "n4a" structure ($I$-3-4,
$V$-2-1) is a hybrid SNS structure.

At temperatures very close to $T_{c}$, the hybridity of the
two-width SNS structure allows observing negative local and
nonlocal voltages (resistances) in the "n4b" structure measured
according to the $I$-3-2, $V$-1-4 circuit at $T=1.460-1.484$ K and
the $I$-3-1, $V$-2-4 circuit at $T=1.455-1.491$ K, respectively
\cite{kuznphysicacnegativ20}.

For the "sl5" structure (Fig. \ref{f3}(a)) in the range of
$T=1.337-1.405$ K, the critical current is fitted to the function
$I_{r}(T)$ (\ref{e4}) with $I_{rf}(0)=28$ $\mu$A and
$T_{r}^{cf}=1.405$ K (Table \ref{t2}). We believe that in this
temperature range, the "sl5" structure is a hybrid NS structure
with critical temperatures in the individual narrow and wide wires
of the structure close to $T_{cn}=1.337$ and $T_{cw}=1.405$ K,
respectively. The linear function $I_{r}(T)$ (\ref{e4}) (solid
line 1b) coincides with the temperature-dependent Josephson
critical current $I_{J}(T)$.

For the "sl6" structure (Fig. \ref{f3}(b)), in the temperature
range of 1.391-1.430 K, the experimental retrapping critical
current depends weakly on temperature and is less than the
switching critical current. In the temperature range of
1.437-1.472 K, the solid line 1b shows the fit of the measured
retrapping critical current to the function $I_{r}(T)$ (\ref{e4}).
We identify this current with the Josephson critical current
$I_{J}(T)$. We assume that at $T=1.437-1.472$ K, the "sl6"
structure is a hybrid NS structure with the critical temperatures
of the narrow and wide wires of the structure close to
$T_{cn}=1.437$ and $T_{cw}=1.472$ K, respectively.

\subsection{Calculated critical currents $I_{GL}(T,B)$ in constant-width structures}

Before showing the measured critical switching and retrapping
currents as functions of the perpendicular magnetic field $B$ at
given temperatures $T_{g}$ in five structures: "n4b", "n4a", "n1",
"sl5", "sl6", we present some theoretical equations needed to fit
the experimental switching current as a function of $T$ and $B$.

At $T$ slightly below $T_{c}$, the GL critical current
$I_{GL}(T,B)$ as a function of temperature $T$ and magnetic field
$B$ parallel to the surface of a thin superconducting film of
thickness $d$ can be calculated by minimizing the Gibbs potential
\cite{tinkham}. In the case of a single quasi-one-dimensional
superconducting wire of constant width $w$ placed in a magnetic
field perpendicular to the substrate surface, it is necessary to
substitute the thickness $d$ for the width $w$ in the equation for
the GL critical current. As a result, we obtain
\begin{equation}
\label{e5} I_{GL}(T,B)=I_{GL}(T,0)(1-B^{2}/B_{c}^{2}(T))^{3/2},
\end{equation}
where $I_{GL}(T,0)= I_{GL}(T)$ is the GL critical current at $T$
in $B=0$ G and $B_{c}(T)=\sqrt{3}\Phi_{0}/(\pi w\xi(T))$ is the
maximum critical field at a given temperature. We rewrite equation
(\ref{e5}) as
\begin{equation}
\label{e6} I_{GL}(T,B)=I_{GL}(T,0)(1-(B\pi
w\xi(T)/(\sqrt{3}\Phi_{0}))^{2})^{3/2}.
\end{equation}

Using the expression for $\xi(T)$ and taking into account that
$I_{GL}(T)=I_{GL}(0)(1-T/T_{c})^{3/2}$, where
$I_{GL}(0)=I_{GL}(T=0, B=0)$ is the Ginzburg-Landau critical
current at $T=0$\,K in $B=0$\,G, we rewrite equation (\ref{e6}) in
a different form
\begin{equation}
\label{e7} I_{GL}(T,B)= I_{GL}(0)(1- \bigg(\frac{B\pi
w\xi(0)}{\sqrt{3}\Phi_{0}}\bigg)^{2}-T/T_{c})^{3/2},
\end{equation}
here $T_{c}=T_{c}(0)$ is the critical temperature at $B=0$\,G.

The equations (\ref{e6}) and (\ref{e7}) are used by us to
calculate the critical switching current, which coincides with the
depairing Ginzburg-Landau critical current $I_{GL}(T,B)$ in a
single wire of constant width.Equation (\ref{e7}) can be written
as:
\begin{equation}
\label{e8}
I_{GL}(T,B)=I_{GL}(0)(1-B^{2}/B_{c}^{2}(0)-T/T_{c})^{3/2},
\end{equation}
where $B_{c}(0)=\sqrt{3}\Phi_{0}/(\pi w\xi(0))$ is the maximum
critical field at $T=0$\,K.

The critical temperature of a quasi-one-dimensional
superconducting wire in a magnetic field normal to the substrate
surface is determined by the equation for the critical temperature
of a thin superconducting film in a parallel magnetic field
\cite{tinkham}
\begin{equation}
\label{e9} T_{c}(B)=T_{c}(1-B^{2}/B_{c}^{2}(0)).
\end{equation}
Then
\begin{equation}
\label{e10} T_{c}(B)=T_{c}(1- (B\pi
w\xi(0)/(\sqrt{3}\Phi_{0}))^{2}).
\end{equation}
Using equation (\ref{e9}), we rewrite equation (\ref{e8}) as
follows:
\begin{equation}
\label{e11} I_{GL}(T,
T_{c}(B))=I_{GL}(0)((T_{c}(B)-T)/T_{c})^{3/2}.
\end{equation}

In the structures we studied, the critical temperature $T_{cw}$ of
the wide wires is higher than the critical temperature $T_{cn}$ of
the narrow wires in the zero magnetic field. It was previously
found by \cite{kuznphysicaccrit22} and confirmed for structures
with different geometries in this work that the switching critical
current as a function of temperature in the zero field in a
quasi-one-dimensional superconducting structure is nonlocal and
depends on the physical parameters of both wide and narrow wires.
The critical magnetic-field-dependent temperatures $T_{cw}(B)$ and
$T_{cn}(B)$ of the wide and narrow wires are determined by
equation (\ref{e10}), in which the corresponding values of $w$ and
$\xi(0)$ are put.

\subsection{Switching $I_{c}(T_{g},B)$ and retrapping $I_{r}(T_{g},B)$ currents vs. $B$ in the "n4b" structure}

\begin{figure}
\begin{center}
\includegraphics[width=1\linewidth]{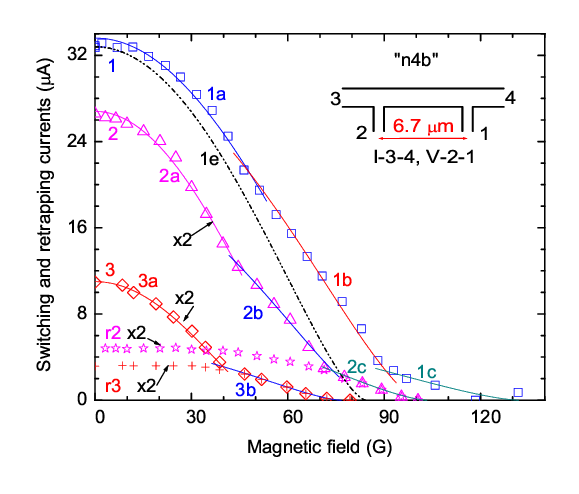}
\caption{\label{f4} (Color online) Experimental data for the "n4b"
structure ($I$-3-4, $V$-2-1): 1 (squares), 2 (triangles), 3
(diamonds) are the switching currents vs. $B$ at temperatures
$T_{g}$; r2 (asterisks), r3 (crosses) are the retrapping currents
vs. $B$ at temperatures corresponding to data 2, 3. For data 2,
r2,3,r3, the scale along the vertical coordinate axis is doubled.
The sets of solid lines: 1a-1b, 2a-2b, 3a, are the fits of the
switching currents vs. $B$ (squares, triangles, diamonds,
respectively), measured in low and intermediate fields, to the
functions $I_{ck}(T_{g},B)$ (\ref{e14}). The solid lines 1c, 2c,
3b (drawn for the eye) indicate the switching currents vs. $B$
(squares, triangles, diamonds, respectively), measured in high
fields.The dash-dotted line 1e is the expected switching current
vs. $B$ at $T_{g}=1.260$ K, calculated using the function
$I_{GL}(T_{g}, T_{c}(B))$ (\ref{e15}).The inset: sketch of the
structure (not to scale) with the measurement circuit.}
\end{center}
\end{figure}

Fig. \ref{f4} shows the measured switching (data 1-3) and
retrapping (data r2, r3) currents as functions of the magnetic
field at different temperatures $T_{g}$ in the "n4b" structure
($I$-3-4, $V$-2-1). Fig. \ref{f4} contains sets of fitted solid
lines 1a-1b, 2a-2b, 3a for data 1-3 and the dash-dotted line 1e
(expected switching current vs. $B$ at $T_{g}=1.260$ K). The solid
lines 1c, 2c, 3b (drawn for the eye) indicate the switching
currents vs. $B$, measured in high fields.

\begin{table*}
\caption{\label{t3} Expected physical and fitted parameters used
to fit the experimental switching current vs. $B$ at temperatures
$T_{g}$ for low and intermediate fields, to the functions
$I_{ck}(T_{g},B)$ (\ref{e14}), where $k=1, 2$ in the "n4b"
structure measured using the circuit $I$-3-4, $V$-2-1 (Fig.
\ref{f4}): data are the experimental data, $T_{g}$ is the
temperature, $\xi(T_{g})_{e}$ is the expected superconducting
coherence length at temperature $T_{g}$, $\xi_{f1}$, $\xi_{f2}$
are the fitted coherence lengths for low and intermediate fields,
$w_{1}$, $ w_{2}$ are the fitted widths that can take only two
values: $w_{k}=0.27$ or 0.48 $\mu$m, $I_{cwe}$ is the expected
switching current calculated for the individual wide wire of the
structure at temperatures $T_{g}$ in the zero field, $I_{cf1}$,
$I_{cf2}$ are the fitted switching currents for low and
intermediate fields.}
\centering %
\begin{tabular}{cccccccccc}
\hline
data & $T_{g}$, & $\xi(T_{g})_{e}$, & $\xi_{f1}$, & $\xi_{f2}$, & $w_{1}$, & $ w_{2}$, & $I_{cwe}$, & $I_{cf1}$, & $I_{cf2}$, \\
  & K & $\mu$m & $\mu$m & $\mu$m & $\mu$m & $\mu$m & $\mu$A & $\mu$A & $\mu$A \\
\hline
a1 & 1.250 & 0.269 & 0.250 & 0.230 & 0.48 & 0.48 & 34.2 & 36.4 & 34.5 \\
a2 & 1.253 & 0.271 & 0.255 & 0.225 & 0.48 & 0.48 & 33.7 & 36.7 & 32.0 \\
1 & 1.260 & 0.275 & 0.255 & 0.235 & 0.48 & 0.48 & 32.3 & 33.6 & 31.0 \\
2 & 1.353 & 0.358 & 0.340 & 0.275 & 0.48 & 0.48 & 14.3 & 13.4 & 10.0 \\
a3 & 1.361 & 0.370 & 0.355 & 0.270 & 0.48 & 0.48 & 12.9 & 13.4 & 9.6 \\
3 & 1.408 & 0.470 & 0.450 &  & 0.48 &  & 5.5    & 5.5   & \\
\hline
\end{tabular}
\end{table*}

We assume that, in a certain limited field range, the measured
switching critical current vs. $B$ at a given temperature $T_{g}$
in the two-width structures we studied, including the "n4b"
structure, can be described by the equation for the GL critical
current $I_{GL}(T,B)$ (\ref{e6}) as a function of temperature $T$
and magnetic field $B$ in a single wire of the constant width $w$,
equal to either the width $w_{n}$ of the narrow wire of the
structure or the width $w_{w}$ of the wide wire of the structure.
Therefore, we fitted the experimental data 1-3 (Fig. \ref{f4}) and
other data to functions defined by the equation
\begin{equation}
\label{e14} I_{ck}(T_{g},B)=I_{cfk}(1-(B\pi
w_{k}\xi_{fk}/(\sqrt{3}\Phi_{0}))^{2})^{3/2},
\end{equation}
where $k$ takes the values of 1, 2 for the first, second functions
corresponding to low, intermediate fields, respectively. In
equation (\ref{e14}), $I_{cfk}=I_{cfk}(T_{g},B=0)$ and
$\xi_{fk}=\xi_{fk}(T_{g})$ are the fitted switching currents and
coherence lengths at temperatures $T_{g}$ in the zero field,
$w_{k}$ are the fitted widths, specially taken equal to either
$w_{n}$ or $w_{w}$. In particular, $w_{n}=0.27$, $w_{w}=0.48$
$\mu$m for "n4b", "n4a" structures. The values of $I_{cfk}$,
$T_{g}$, $\xi_{fk}$, $w_{k}$ and other values are given in Table
\ref{t3}.

We found that the switching current as a function of magnetic
field is nonlocal and depends on the physical parameters of the
area containing the junction line of the narrow and wide wires of
the structure. The choice of the fitted width $w_{n}$ or $w_{w}$
indicates that, within a certain limited range of fields, the
switching current is determined primarily by the physical
parameters of the narrow or wide wire.

For different magnetic field ranges, the experimental switching
current (data a1, a2, a3 (Table \ref{t3}), not shown in the
figure, and data 1-3 (Fig. \ref{f4})) is fitted to the functions
$I_{ck}(T_{g},B)$ (\ref{e14}). Table \ref{t3} shows that for these
data, the fitted coherence length $\xi_{f1}$ corresponding to low
fields is close to the expected coherence length $\xi(T_{g})_{e}$
at temperatures $T_{g}$ in the zero field. When calculating
$\xi(T_{g})_{e}$, we took the critical temperature $T_{c}(0.5)$
(Tables \ref{t1}, and \ref{t2}). The value of $\xi_{f2}$ is
shorter than the value of $\xi_{f1}$.

Fig. \ref{f4} shows that the steepness of lines 1b, 2b, 3b is less
than the steepness of the corresponding lines 1a, 2a, 3a. The
steepness of lines 1c and 2c is clearly less than the steepness of
lines 1a, 2a. We found that the fields corresponding to the
intersections of lines 1b and 1c, lines 2b and 2c, and the middle
of line 3b are practically equal to the fields at which the
individual (without taking into account the influence of the
proximity effect) wide wire of the structure should transition to
the normal state. These fields are calculated from the condition
$T_{c}(B)=T_{g}$, where $T_{c}(B)$ is given by equation
(\ref{e10}), into which the corresponding values $w=w_{w}=0.48$
$\mu$m, $\xi(0)=0.107$ $\mu$m, $T_{c}=T_{r}^{cf}=1.510$ K are
substituted (Tables \ref{t1}, and \ref{t2}). A slight decrease in
the steepness (lines 1b, 2b, 3b) is due to the influence of narrow
potential wires of the structure, in which superconductivity is
kept in high fields.

In Table \ref{t3}, we have placed the expected values of the
switching current $I_{cwe}=I_{cwe}(T_{g},B=0)$ of the individual
wide wire of the "n4b" structure at temperatures $T_{g}$ in the
zero field, calculated using the KL fitting function $I_{KL}(T)$
(\ref{e3}) obtained in \cite{kuznphysicaccrit22} with two fitted
parameters placed in Table \ref{t2}. The measured switching
current $I_{cwm}=I_{cwm}(T_{g},B=0)$ at temperatures $T_{g}$ in
the zero field (Fig. \ref{f4}) is close to the value
$I_{cwe}=I_{cwe}(T_{g},B=0)$ and practically coincides with the
fitted switching current $I_{cf1}$ corresponding to low fields.

We calculated the expected switching current as a function of $B$
at a given temperature $T_{g}$ in the two-width "n4b" ($I$-3-4,
$V$-2-1) structure, neglecting the influence of narrow potential
wires of the structure on electron transport and using the
equation for the GL critical current $I_{GL}(T,B)$ (\ref{e7}) as a
function of temperature $T$ and magnetic field $B$ in a single
wire of the constant width $w$, equal to the width $w_{w}$ of the
wide wire of the structure. For this purpose, we replaced the
values of $T$, $I_{GL}(0)$, $T_{c}$ in equation (\ref{e7}) with
the given temperature $T_{g}$ and the fitted values of
$I_{GL}^{f}(0)$, $T_{GL}^{cf}$ (Table \ref{t2}), found earlier in
\cite{kuznphysicaccrit22}. As a result, we obtained the equation
(\ref{e15})
\begin{equation}
\label{e15} I_{GL}(T_{g}, T_{c}(B))=I_{GL}^{f}(0)(1-
\bigg(\frac{B\pi
w\xi(0)}{\sqrt{3}\Phi_{0}}\bigg)^{2}-T_{g}/T_{GL}^{cf})^{3/2}.
\end{equation}
In equation (\ref{e15}), the width $w$ takes the values of $w_{n}$
and $w_{w}$ for the narrow and wide wires of the structure,
respectively.

The dash-dotted line (1e, Fig. \ref{f4}) shows the expected
switching current vs. $B$ at $T_{g}=1.260$ K in the individual
wide wire of the structure. Line 1e is plotted using equation
(\ref{e15}) with the corresponding parameters from Tables
\ref{t1}, and \ref{t2}. It can be seen that at high fields, line
1e deviates significantly from the measured switching current as a
function of $B$ at $T_{g}=1.260$ K (data 1 - squares, Fig.
\ref{f4}). Moreover, line 1e crosses the field coordinate axis at
$B=84$ G. However, the experimental switching current (data 1)
exists up to $B=130$ G. We believe that at high fields, the
superconducting order parameter in the structure is kept due to
the proximity effect near the junction line of the wide and narrow
wires. Indeed, we found that at $T_{g}=1.260$ K, the individual
narrow wire of the structure (without taking into account the
influence of the wide wire of the structure) should transition to
the normal state in the field $B=141.0$ G. The value of $B=141.0$
G is obtained from the condition $T_{g}=T_{c}(B)$. The critical
temperature $T_{c}(B)$ (\ref{e10}) is calculated for the
individual narrow wire of the structure using the corresponding
physical parameters, including $T_{c}=T_{GL}^{cf}=1.437$ K (Tables
\ref{t1}, and \ref{t2}).

The solid lines 1c, 2c, 3b show areas where there is the low
superconducting critical current that is weakly dependent on the
magnetic field. At the same time, the switching and retrapping
currents coincide and the individual wide current wire of the
structure must be in the normal state. The non-zero critical
current in fields larger than the maximum critical field is kept
due to the proximity effect and equilibrium fluctuations of the
superconducting order parameter \cite{kuznphysicacnegativ20} in
the narrow potential wire of the structure.

\begin{figure}
\begin{center}
\includegraphics[width=1\linewidth]{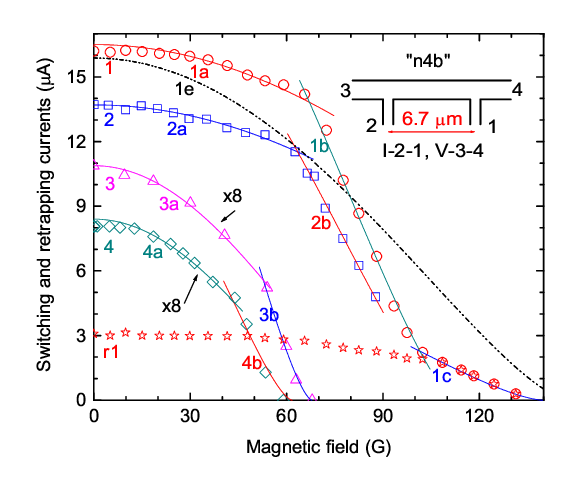}
\caption{\label{f5} (Color online) Experimental data for the "n4b"
structure ($I$-2-1, $V$-3-4): 1 (circles), 2 (squares), 3
(triangles), 4 (diamonds) are switching currents vs. $B$ at
temperatures $T_{g}$; r1 (asterisks) is the retrapping current vs.
$B$ at $T_{g}=1.243$ K. For data 3 and 4, the switching and
retrapping currents are the same. For data 3, 4, the vertical
coordinate axis scale is increased eightfold. The sets of solid
lines: 1a-1b, 2a-2b, 3a, 4a are the fits of the switching currents
vs. $B$ (circles, squares, triangles, diamonds, respectively),
measured in low and intermediate fields, to the functions
$I_{ck}(T_{g},B)$ (\ref{e14}). The solid lines 1c, 3b, 4b (drawn
for the eye) indicate the switching currents vs. $B$ (circles,
triangles, diamonds, respectively), measured in high fields. The
dash-dotted line 1e shows the expected switching current vs. $B$
at $T_{g}=1.243$ K, calculated using the function $I_{GL}(T_{g},
T_{c}(B))$ (\ref{e15}). The inset: sketch of the structure (not to
scale).}
\end{center}
\end{figure}

\begin{table*}
\caption{\label{t4} Expected and fitted parameters used to fit the
experimental switching current vs. $B$ at temperatures $T_{g}$ for
two field ranges to the functions $I_{ck}(T_{g},B)$ (\ref{e14}),
where $k=1, 2$ in the "n4b" structure measured according to the
measurement circuit $I$-2-1, $V$-3-4 (Fig. \ref{f5}). The notation
$I_{cne}$ is the expected switching current calculated for the
narrow wire of the structure at temperatures $T_{g}$ in the zero
field. Explanations of other notations are given in the title of
Table \ref{t3}.}
\centering %
\begin{tabular}{ccccccccccc}
\hline
data & $T_{g}$, & $\xi(T_{g})_{e}$, & $\xi_{f1}$, & $\xi_{f2}$, & $w_{1}$, & $ w_{2}$, & $I_{cwe}$, & $I_{cne}$, & $I_{cf1}$, &$I_{cf2}$, \\
  & K & $\mu$m  &  $\mu$m &  $\mu$m &  $\mu$m &  $\mu$m &  $\mu$A &  $\mu$A & $\mu$A & $\mu$A \\
\hline
1 & 1.243 & 0.260 & 0.21 & 0.210 & 0.27 & 0.48 & 35.8 & 16.1 & 16.5 & 26.5 \\
a1 & 1.254 & 0.266 & 0.21 & 0.205 & 0.27 & 0.48 & 33.4 & 14.7 & 15.5 & 23.0 \\
2 & 1.267 & 0.273 & 0.22 & 0.215 & 0.27 & 0.48 & 30.8 & 13.3 & 13.7 & 21.0  \\
3 & 1.410 & 0.473 & 0.48 & & 0.27 & &          5.2 & 1.3 & 1.36 &   \\
a2 & 1.422 & 0.515 & 0.55 & & 0.27 &  &      3.7 & 1.2 & 1.19 &   \\
4 & 1.435 & 0.576 & 0.56 &   & 0.27 &  &     2.2 & 1.0 & 1.05 &   \\
\hline
\end{tabular}
\end{table*}

Fig. \ref{f5} shows that the steepness of lines 1b, 2b, 3b, 4b are
greater than the steepness of the corresponding lines 1a, 2a, 3a,
4a. The steepness of line 1c is less than the steepness of line
1a. We found that the fields corresponding to the intersection of
lines 1b and 1c and the midpoints of lines 3b, 4b are close to the
fields at which the individual wide wire of the structure should
transition to the normal state. These fields are found from the
condition $T_{c}(B)=T_{g}$, where $T_{c}(B)$ is given by equation
(\ref{e10}), into which the necessary physical parameters are
placed, including $T_{c}=T_{r}^{cf}=1.525$ K (Tables \ref{t1}, and
\ref{t2}).

We found that the experimental switching current vs. $B$ is
determined by a limited area, containing the junction line of the
narrow and wide current wires. In low fields, the switching
current vs. $B$ is determined primarily by electron transport in
the narrow wire. We assume that in a narrow intermediate field
range, the sharp decrease in the switching current by a large
magnitude with increasing field (large steepness of lines 1b, 2b)
is due to the fact that the switching current is determined
primarily by electron transport in the wide current wire.

In high fields, the solid lines 1c, 3b, 4b demonstrate a slight
decrease in the small switching current with increasing field. In
high fields, the switching and retrapping currents coincide and
the individual wide wire of the structure must be in the normal
state. The non-zero critical current in fields larger than the
maximum critical field is kept due to the proximity effect near
the junction line of the wide and narrow wires and equilibrium
fluctuations of the superconducting order parameter.

We calculated the expected switching current vs. $B$ at
$T_{g}=1.243$ K in the two-width "n4b" ($I$-2-1, $V$-3-4)
structure, neglecting the influence of wide wires of the structure
on electron transport and using the equation for the GL critical
current in the individual narrow wire $I_{GL}(T_{g}, T_{c}(B))$
(\ref{e15}), into which the corresponding parameters from Tables
\ref{t1}, and \ref{t2} are inserted (dash-dotted line 1e, Fig.
\ref{f5}). The measured switching current vs. $B$ at $T_{g}=1.243$
K in the "n4b" structure (data1- circles, Fig. \ref{f5}) differs
from the expected switching current. The expected switching
current becomes zero in the field $B=147.7$ G.

From Table \ref{t4} it is seen that the ratios
$\xi(T_{g})_{e}/\xi_{f1}$ are 1.24-1.26 and 0.94-1.03 for data
1,a1,2 and 3,a2,4, respectively. When calculating
$\xi(T_{g})_{e}$, we took the critical temperature $T_{c}(0.5)$
(Tables \ref{t1}, and \ref{t2}). For data 1,a1,2, the value of
$\xi_{f2}$ is almost equal to the value of $\xi_{f1}$.

In Table \ref{t4}, the expected values of the switching current
$I_{cwe}=I_{cwe}(T_{g},B=0)$ of the individual wide wire of the
"n4b" structure at temperatures $T_{g}$ lying in the range of
1.250-1.466 K in the zero field are determined in the same way as
these values are calculated in Table \ref{t3}.

In Table \ref{t4}, the expected values of the switching current
$I_{cne}=I_{cne}(T_{g},B=0)$ of the individual narrow wire of the
"n4b" structure at temperatures $T_{g}$ lying in the range of
1.250-1.400 K in the zero field were calculated using the K-L
fitting function $I_{KL}(T)$ (\ref{e3}) obtained in
\cite{kuznphysicaccrit22} with two fitted parameters placed in
Table \ref{t2}.

In the range of 1.410-1.475 K, the temperature-dependent switching
and retrapping currents measured in the zero field in the "n4b"
structure ($I$-2-1, $V$-3-4) coincide \cite{kuznphysicaccrit22}.
Therefore, to calculate $I_{cne}=I_{cne}(T_{g},B=0)$ at
temperatures $T_{g}$ corresponding to data 3,a2,4, we took the
equation $I_{r}(T)=I_{rf}(0)(1-T/T_{r}^{cf})$ (\ref{e4}), where
the fitted retrapping current $I_{rf}(0)=17$ $\mu$A at $T=0$ K and
the fitted critical temperature $T_{r}^{cf}=1.525$ K, greater than
$T_{c}(0.5)=1.487$ K (Table \ref{t2}).

The measured switching current $I_{cnm}=I_{cnm}(T_{g},B=0)$ at
temperatures $T_{g}$ in the zero field (data 1-4 taken at $B=0$ G,
Fig. \ref{f5}) practically coincides with the fitted switching
current $I_{cf1}$ and is close to $I_{cne}=I_{cne}(T_{g},B=0)$.
The value of $I_{cf2}$ is greater than the value of $I_{cf1}$
(Table \ref{t4}).

In the intermediate field range, the greater influence of the wide
current wire on the switching current vs. $B$ is confirmed by the
fact that the equation for $I_{c2}(T_{g},B)$ (\ref{e14}) contains
the fitted width $w_{2}=w_{w}=0.48$ $\mu$m and the fitted critical
current $I_{cf2}$ satisfying the relations
$I_{cwe}/I_{cf2}=1.35-1.86$ and $\sqrt{I_{cwe}I_{cne}} \approx
I_{cf2}$ (Table \ref{t4}).

In addition, we used the $I$-3-1, $V$-2-4 measurement circuit to
record the $I$-$V$ curves in the "n4b" structure located in a
magnetic field at given temperatures. For the $I$-2-1, $V$-3-4
(Fig. \ref{f5}) and $I$-3-1, $V$-2-4 circuits, practical agreement
was found between the magnetic-field dependences of the critical
currents measured at the same temperature, not very close to
$T_{c}$. The measurement results according to the $I$-3-1, $V$-2-4
circuit are not presented here.

\subsection{Switching $I_{c}(T_{g},B)$ and retrapping $I_{r}(T_{g},B)$ currents vs. $B$ in the "n4a" structure}

\begin{figure}
\begin{center}
\includegraphics[width=1\linewidth]{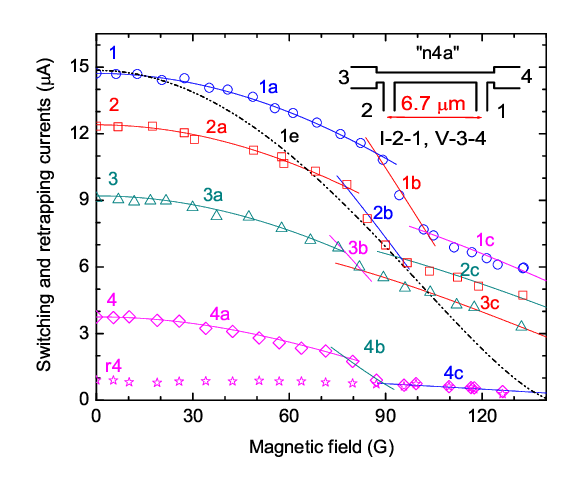}
\caption{\label{f6} (Color online) Experimental data for the "n4a"
structure ($I$-2-1, $V$-3-4): 1 (circles), 2 (squares), 3
(triangles), 4 (diamonds) are switching currents vs. $B$ at
temperatures $T_{g}$; r4 (asterisks) is the retrapping current vs.
$B$ at 1.361 K. The four solid lines: 1a, 2a, 3a, 4a are the fits
of switching currents vs. $B$ (circles, squares, triangles,
diamonds, respectively), measured in low fields, to the function
$I_{c1}(T_{g},B)$ (\ref{e14}). The four pairs of solid lines: 1b
and 1c, 2b and 2c, 3b and 3c, 4b and 4c (drawn for the eye)
indicate the switching currents vs. $B$ (circles, squares,
triangles, diamonds, respectively), measured in the intermediate
and high field ranges, respectively. The dash-dotted line 1e is
the expected switching current vs. $B$ at 1.257 K, calculated
using the function $I_{GL}(T_{g}, T_{c}(B))$ (\ref{e15}). The
inset: sketch of the structure (not to scale).}
\end{center}
\end{figure}

Fig. \ref{f6} shows the switching current as a function of
magnetic field measured in the n4a structure using the $I$-2-1,
$V$-3-4 circuit at temperatures $T_{g}$ (data 1 - circles, 2 -
squares, 3 - triangles, 4 - diamonds). The experimental retrapping
current r4 (asterisks) is shown at 1.361 K. Solid lines: 1a, 2a,
3a, 4a are the fits of the switching currents vs. $B$ (data 1-4),
measured at low fields, to the function $I_{c1}(T_{g},B)$
(\ref{e14}). The solid lines: 1b and 1c, 2b and 2c, 3b and 3c, 4b
and 4c (drawn for the eye) indicate the switching currents vs. $B$
(data 1-4), measured in the intermediate and high field ranges,
respectively.

The steepness of lines 1b, 2b, 3b, 4b are greater than the
steepness of the corresponding lines 1a, 2a, 3a, 4a. In the
intermediate field range, lines 1b, 2b, 3b, 4b indicate the
influence of the physical parameters of the wide potential wire on
a sharp decrease in the switching current by a small magnitude
with increasing field. We found that the fields corresponding to
the midpoints of lines 1b, 2b, the intersection of lines 3b and
3c, the intersection of lines 4a and 4b are close to the fields at
which the individual wide wire of the structure should transition
to the normal state. The steepness of lines 1c, 2c, 3c, 4c are
less than the steepness of the corresponding lines 1a, 2a, 3a, 4a.
Lines 1c, 2c, 3c, 4c demonstrate that the switching current
slightly decreases with increasing field and is kept in high
fields.

In high fields, the switching and retrapping currents coincide and
the individual wide wire of the structure must be in the normal
state. The critical current in high fields is kept due to the
proximity effect near the junction line of the wide and narrow
wires and equilibrium fluctuations of the superconducting order
parameter.

We plotted the expected switching current vs. $B$ at $T_{g}=1.257$
K in the two-width "n4a" ($I$-2-1, $V$-3-4) structure by
neglecting the influence of wide wires of the structure on
electron transport and using the equation for the GL critical
current in the individual narrow wire $I_{GL}(T_{g}, T_{c}(B))$
(\ref{e15}), into which the corresponding parameters from Tables
\ref{t1}, and \ref{t2} are inserted (dash-dotted line 1e, Fig.
\ref{f6}). The experimental switching current vs. $B$ at
$T_{g}=1.257$ K in the "n4a" structure (data 1-circles, Fig.
\ref{f6}) differs significantly from the expected switching
current. Line 1e intersects the field axis at the value of
$B=142.1$ G, at which the individual narrow wire of the structure
should transition to the normal state. However, the measured
switching current vs. $B$ (data 1, Fig. \ref{f6}) is kept in the
field of 142.1 G.

In low fields, the switching current vs. $B$ (1-4, Fig. \ref{f6}),
measured at temperatures $T_{g}$, is fitted to the functions
(solid lines 1a, 2a, 3a, 4a) $I_{ck}(T_{g},B)$ (\ref{e14}), where
$k=1$. From Table \ref{t5} it is seen that
$\xi(T_{g})_{e}>\xi_{f1}$.

\begin{table*}
\caption{\label{t5} Expected and fitted parameters used to fit the
experimental switching current vs. $B$ at temperatures $T_{g}$ for
low fields to the functions $I_{ck}(T_{g},B)$ (\ref{e14}), where
$k=1$ in the "n4a" structure measured according to the $I$-2-1,
$V$-3-4 circuit (Fig. \ref{f6}). The meanings of the notations:
$T_{g}$, $\xi(T_{g})_{e}$ and others are explained in the title of
Table \ref{t3}.}
\centering %
\begin{tabular}{cccccccc}
\hline
data & $T_{g}$, & $\xi(T_{g})_{e}$, & $\xi_{f1}$, & $\xi(T_{g})_{e}/\xi_{f1}$ & $w_{1}$,  & $I_{cne}$, &  $I_{cf1}$,  \\
 & K & $\mu$m & $\mu$m &  &  $\mu$m &  $\mu$A & $\mu$A \\
\hline
1 & 1.257 & 0.280 & 0.200 & 1.40 & 0.27 & 14.5 & 14.72 \\
2 & 1.273 & 0.295 & 0.215 & 1.37 & 0.27 & 12.6 & 12.40 \\
3 & 1.300 & 0.320 & 0.240 & 1.33 & 0.27 &  9.7  & 9.20  \\
4 & 1.361 & 0.410 & 0.325 & 1.25 & 0.27 &  3.8  & 3.75 \\
\hline
\end{tabular}
\end{table*}
Table \ref{t5} shows the expected values of the switching current
$I_{cne}=I_{cne}(T_{g},B=0)$ of the individual narrow wire of the
"n4a" structure ($I$-2-1, $V$-3-4) at temperatures $T_{g}$ in the
zero field, which are calculated using the K-L fitting function
$I_{KL}(T)$ (\ref{e3}) with two corresponding fitted parameters
listed in Table \ref{t2}. The measured switching current
$I_{cnm}=I_{cnm}(T_{g},B=0)$ at temperatures $T_{g}$ in $B=0$ G
(Fig. \ref{f6}) practically coincides with the fitted switching
current $I_{cf1}$ and is close to $I_{cne}=I_{cne}(T_{g},B=0)$
(Table \ref{t5}).

\begin{figure}
\begin{center}
\includegraphics[width=1\linewidth]{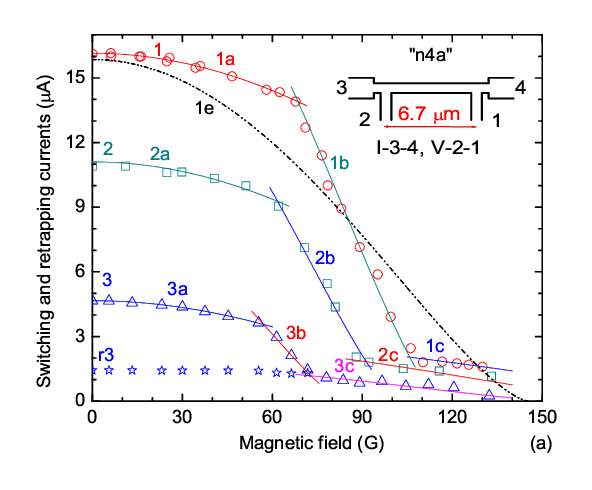}
\includegraphics[width=1\linewidth]{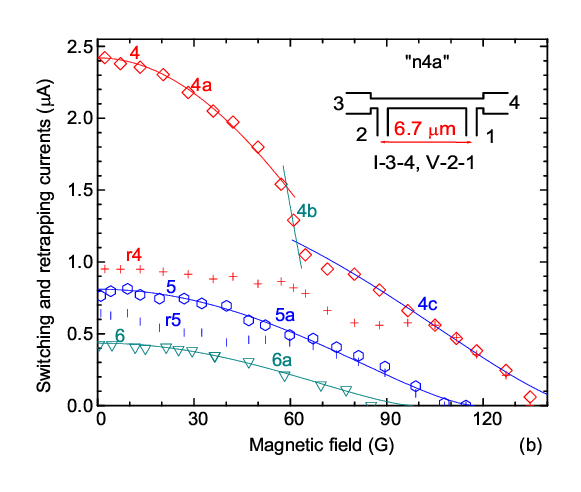}
\caption{\label{f7} (Color online) (a)-(b) Experimental data for
the "n4a" structure \, ($I$-3-4, $V$-2-1): (a) 1 (circles), 2
(squares), 3 (up triangles) and (b) 4 (diamonds), 5 (hexagons), 6
(down triangles) are the switching currents vs. $B$ at
temperatures $T_{g}$; (a) r3 (asterisks), and (b) r4 (crosses), r5
(vertical bars) are the retrapping currents vs. $B$ at
temperatures corresponding to data 3,4,5. The sets of solid lines:
(a) 1a-1b, 2a-2b, 3a-3b, and (b) 4a-4b are the fits of the
measured switching currents vs. $B$ ((a) circles, squares,
triangles, and (b) diamonds, respectively) to the functions
$I_{ck}(T_{g},B)$ (\ref{e14}) in  low and intermediate fields.
Solid lines: (a) 1c, 2c, 3c, and (b) 4c (built for the eye)
indicate the measured switching currents vs. $B$ ((a) circles,
squares, triangles, and (b) diamonds, respectively) in high
fields. (b) Solid lines 5a, 6a (plotted for the eye) indicate the
measured switching currents vs. $B$ (hexagons, down triangles,
respectively) at temperatures of 1.404, 1.414 K, respectively, in
all field ranges. (a) Dash-dotted line 1e shows the expected
switching current vs. $B$ at the temperature corresponding to data
1, calculated using the function $I_{GL}(T_{g}, T_{c}(B))$
(\ref{e15}). (a)-(b) The inset: sketch of the structure (not to
scale) with the measurement circuit.}
\end{center}
\end{figure}

Fig. \ref{f7}(a)-(b) contains the switching current as a function
of magnetic field at different temperatures measured in the "n4a"
structure using the $I$-3-4, $V$-2-1 circuit: (a) data 1- circles,
2 - squares, 3 - up triangles, and (b) 4 - diamonds, 5 - hexagons,
6 - down triangles.  In this figure, (a) data r3 - asterisks, (b)
r4 - crosses, r5 -vertical bars are retrapping currents vs. $B$,
measured at temperatures of 1.345, 1.379, 1.404 K, respectively.
At a temperature of 1.414 K, the switching and rettraping currents
coincide.

The sets of solid lines: (a) 1a-1b, 2a-2b, 3a-3b, and (b) 4a-4b
are the fits of the measured switching currents vs. $B$ (data 1-4,
respectively) to the functions $I_{ck}(T_{g},B)$ (\ref{e14}) in
low and intermediate fields. Solid lines: (a) 1c, 2c, 3c, and (b)
4c (drawn for the eye) indicate the measured switching currents
vs. B ((a) circles, squares, triangles, and (b) diamonds,
respectively) in high fields. (b) Solid lines 5a, 6a (drawn for
the eye) indicate the measured switching currents vs. B (data 5-6,
respectively) in all field ranges. Dash-dotted line 1e shows the
expected switching current vs. $B$.

The steepness of lines 1b, 2b, 3b, 4b are greater than the
steepness of the corresponding lines 1a, 2a, 3a, 4a. In the
intermediate fields, lines 1b, 2b, 3b, 4b indicate a strong
influence of the physical parameters of the wide current wire on
the sharp decrease in the switching current by a large magnitude
with increasing field. Note that the height of the switching
current jump in the "n4a" structure ($I$-3-4, $V$-2-1) is greater
than the height of the jump in the "n4a" structure ($I$-2-1,
$V$-3-4). We found that the fields corresponding to the middle of
line 1b, the intersections of lines 2b and 2c, 3b and 3c, 4b and
4c are close to the fields at which the individual wide wire of
the structure should transition to the normal state. The steepness
of lines 1c, 2c, 3c, 4c are smaller than the steepness of the
corresponding lines 1a, 2a, 3a, 4a. At high fields, lines 1c, 2c,
3c, 4c show a small switching current, which decreases slightly
with increasing field. Lines 5a, 6a show areas where the switching
and retrapping currents, measured at temperatures very close to
the critical temperature, coincide over the entire range of
fields.

We calculated the expected switching current vs. $B$ at
$T_{g}=1.249$ K in the two-width "n4a" ($I$-3-4, $V$-2-1)
structure, neglecting the influence of wide current wires of the
structure on electron transport and using the equation for the GL
critical current in the individual narrow wire $I_{GL}(T_{g},
T_{c}(B))$ (\ref{e15}), into which the corresponding parameters
from Tables \ref{t1}, and \ref{t2} are substituted (dash-dotted
line 1e, Fig. \ref{f7}(a)). It is seen that the experimental
switching current vs. $B$ at $T_{g}=1.249$ K in the "n4a"
structure (data 1-circle, Fig. \ref{f7}(a)) differs from the
expected switching current, which disappears in a field $B=145.8$
G. At this field value, the individual narrow wire of the
structure should transition to the normal state. The critical
current in high fields (the solid lines 1c, 2c, 3c, 4c) is kept
due to the proximity effect near the junction line of the wide and
narrow wires and equilibrium fluctuations of the superconducting
order parameter.

Table \ref{t6} shows the physical and fitted parameters used to
fit the switching critical current as a function of magnetic field
measured in the "n4a" structure at temperatures $T_{g}$ (data 1-4,
Fig. \ref{f7}(a) -(b) and data a1-a3, not shown in the figures).

\begin{table*}
\caption{\label{t6} Expected physical and fitted parameters used
to fit the experimental switching current vs. $B$ at temperatures
$T_{g}$ to the functions $I_{ck}(T_{g},B)$ (\ref{e14}), where
$k=1, 2$ for two field ranges in the "n4a" structure measured
according to the $I$-3-4, $V$-2-1 circuit (Fig. \ref{f7}(a)-(b) ).
$I_{cne}$ is the expected switching current calculated for the
individual narrow wire of the "n4a" structure at temperatures
$T_{g}$ in the zero field. The meanings of other notations are
given in the title of Table \ref{t3}.}
\centering %
\begin{tabular}{ccccccccccc}
\hline
data & $T_{g}$, & $\xi(T_{g})_{e}$, & $\xi_{f1}$,  & $\xi_{f2}$, & $w_{1}$, & $w_{2}$, & $I_{cwe}$, & $I_{cne}$, & $I_{cf1}$, & $I_{cf2}$, \\
 & K & $\mu$m & $\mu$m &  $\mu$m &  $\mu$m &  $\mu$m &  $\mu$A &  $\mu$A & $\mu$A &  $\mu$A \\
\hline
1 & 1.249 & 0.28 & 0.19 & 0.20 & 0.27 & 0.48 & 34.4 & 15.3 & 16.14 & 25.00 \\
a1 & 1.260 & 0.29 & 0.20 & 0.21 & 0.27 & 0.48  & 32.2 & 14.1 & 14.70 & 22.00 \\
2 & 1.284 & 0.31 & 0.23 & 0.23 & 0.27 & 0.48 & 27.3 & 11.4 & 11.10 & 18.00  \\
a2 & 1.300 & 0.32 & 0.23 & 0.23 & 0.27 & 0.48 & 24.2 & 9.7 & 9.17 & 14.20 \\
3 & 1.345 & 0.38 & 0.30 & 0.28 & 0.27 & 0.48 & 15.7 & 5.3 & 4.66 & 8.80 \\
a3 & 1.360 & 0.41 & 0.30  & 0.29 & 0.27 & 0.48 & 13.0 & 3.9 & 3.60 & 6.20 \\
4 & 1.379 & 0.46 & 0.37 & 0.32  & 0.27 & 0.48 & 9.8 & 2.4 & 2.42 & 6.00  \\
\hline
\end{tabular}
\end{table*}

For different field ranges, the switching current vs. $B$ at
temperatures $T_{g}$ (data 1-4, Fig. \ref{f7}(a)-(b) and a1-a3
(not shown in the figure)), we fitted to the functions
$I_{ck}(T_{g},B)$ (\ref{e14}). From Table \ref{t6} it is seen that
$\xi(T_{g})_{e}>\xi_{f1}$. For data 1-3 and a1-a3, the fitted
coherence lengths $\xi_{f1}$, $\xi_{f2}$ are very close to each
other. For data 4, $\xi_{f2}<\xi_{f1}$.

In Table \ref{t6}, we calculated the expected values of the
switching current $I_{cwe}$ of the individual wide wire of the
"n4a" structure at temperatures $T_{g}$ in the zero field using
the K-L fitting function $I_{KL}(T)$ (\ref{e3}) for the individual
wide wire of the "n4b" structure ($I$-3-4, $V$-2-1) with the two
corresponding fitted parameters placed in Table \ref{t2}.

In Table \ref{t6}, the expected values of the switching current
$I_{cne}$ of the individual narrow wire of the "n4a" structure
($I$-3-4, $V$-2-1) at temperatures $T_{g}$, in the range of
1.250-1.410 K, in $B=0$ G, are calculated using the K-L fitting
function $I_{KL}(T)$ (\ref{e3}) with two corresponding fitted
parameters placed in Table \ref{t2}.

The measured switching current $I_{cnm}=I_{cnm}(T_{g},B=0)$ (data
1-6 in Fig. \ref{f7}(a)-(b), taken in the zero magnetic field) is
close to $I_{cne}$ and practically coincides with the fitted
critical current $I_{cf1}$. The value of $I_{cf2}$ is
significantly greater than the value of $I_{cf1}$. The ratio
$I_{cwe}/I_{cf2}=1.38-2.15$ indicates a large influence of the
physical parameters of the wide current wires of the "n4a"
structure ($I$-3-4, $V$-2-1) on the switching current vs. $B$
measured in intermediate fields at temperatures $T_{g}$ (Table
\ref{t6}).

Additionally, the $I$-$V$ curves were measured using the $I$-3-1,
$V$-2-4 circuit in the "n4a" structure placed in a magnetic field
at given temperatures. For the $I$-3-4, $V$-2-1 (Fig.
\ref{f7}(a)-(b)) and $I$-3-1, $V$-2-4 circuits, we found
near-identical magnetic-field dependences of the critical currents
recorded at the same temperatures. The results of these
measurements are not shown here.

\subsection{Switching $I_{c}(T_{g},B)$ and retrapping $I_{r}(T_{g},B)$ currents in structures "n1", "sl5", "sl6"}

\begin{figure}
\begin{center}
\includegraphics[width=1\linewidth]{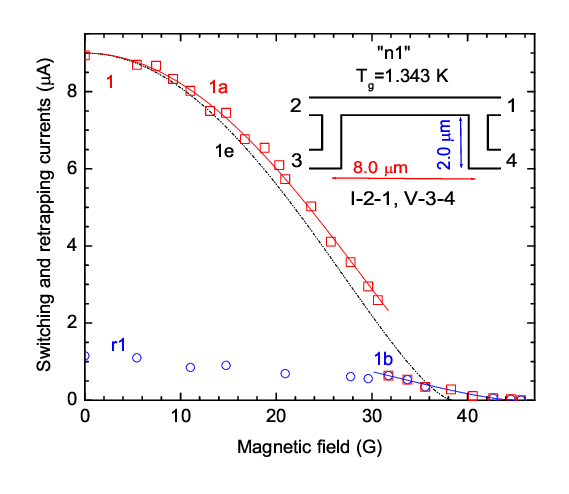}
\caption{\label{f8} (Color online) Experimental data for the "n1"
structure ($I$-2-1, $V$-3-4): 1 (squares) and r1 (circles) are the
switching and retrapping currents vs. $B$ at $T_{g}=1.343$ K,
respectively. Solid line 1a is the fit of the switching current
vs. $B$ (squares), measured in low fields, to the function
$I_{c1}(T_{g},B)$ (\ref{e14}). Solid line 1b (drawn for the eye)
indicates the measured switching and retrapping currents vs. $B$
(squares and circles, respectively), in high fields.  Dash-dotted
line 1e is the expected switching current vs. $B$ at $T_{g}=1.343$
K, calculated using the function $I_{GL}(T,B)$ (\ref{e6}). The
inset: sketch of the structure (not to scale).}
\end{center}
\end{figure}

\begin{figure}
\begin{center}
\includegraphics[width=1\linewidth]{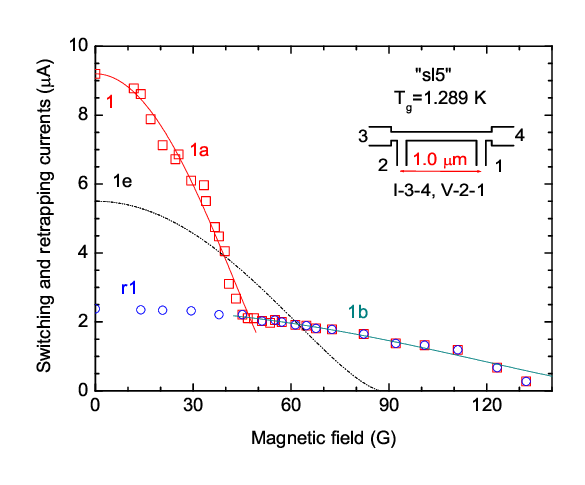}
\caption{\label{f9} (Color online) Experimental data for the "sl5"
structure ($I$-3-4, $V$-2-1): 1 (squares) and r1 (circles) are the
switching and retrapping currents vs. $B$ at $T_{g}=1.289$ K,
respectively. Solid line 1a is the fit of the switching current
vs. $B$ (squares), measured in low fields, to the function
$I_{c1}(T_{g},B)$ (\ref{e14}). Solid line 1b (drawn for the eye)
indicates the measured switching and retrapping currents vs. $B$
(squares and circles, respectively), in high fields. Dash-dotted
line 1e is the expected switching current vs. $B$ at $T_{g}=1.289$
K, calculated using the function $I_{GL}(T, B)$ (\ref{e7}). The
inset: sketch of the structure (not to scale).}
\end{center}
\end{figure}

\begin{figure}
\begin{center}
\includegraphics[width=1\linewidth]{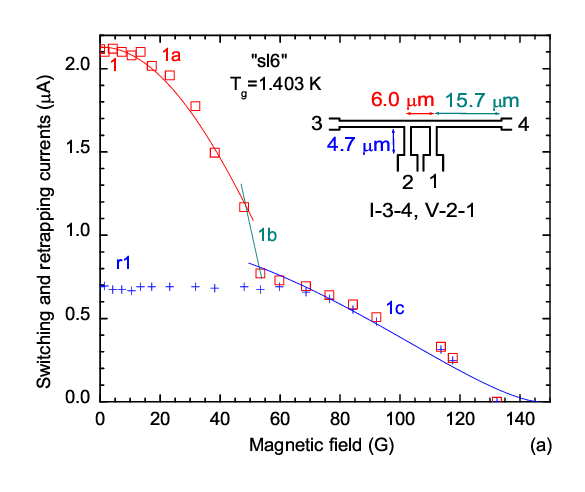}
\includegraphics[width=1\linewidth]{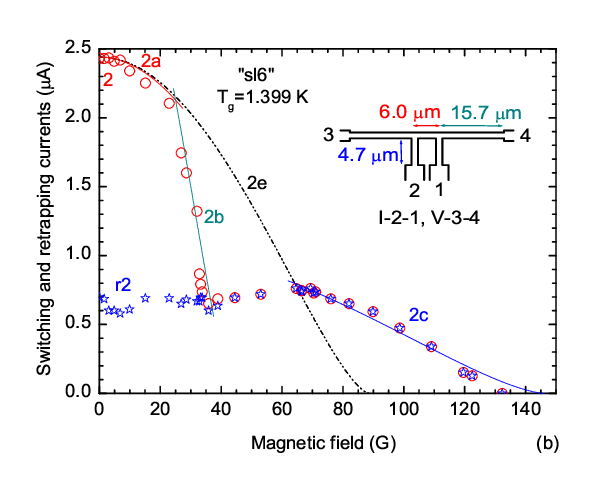}
\caption{\label{f10} (Color online) (a)-(b) Experimental data for
the "sl6" structure (a) ($I$-3-4, $V$-2-1) and (b) ($I$-2-1,
$V$-3-4): (a) 1 (squares) and (b) 2 (circles) are the switching
currents vs. $B$ at $T_{g}=1.403$ and 1.399 K, respectively; (a)
r1 (crosses) and (b) r2 (asterisks) are the retrapping currents
vs. $B$ at 1.403 and 1.399 K, respectively. The sets of solid
lines: (a) 1a-1b, and (b) 2a-2b are the fits of the measured
switching currents vs. $B$ ((a) squares and (b) circles,
respectively) to the functions $I_{ck}(T_{g},B)$ (\ref{e14}),
where $k=1, 2$ in low, intermediate fields. Solid lines: (a) 1c,
and (b) 2c (drawn for the eye) indicate the measured switching
currents vs. $B$ ((a) squares and (b) circles, respectively), in
high fields. (b) The dash-dotted line 2e is the expected switching
current vs. $B$ at $T_{g}=1.399$ K, calculated using the function
$I_{GL}(T,B)$ (\ref{e6}). (a)-(b) The insets: sketches of the
structure (not to scale).}
\end{center}
\end{figure}

Table \ref{t7} shows the physical and fitted parameters used to
fit the switching critical current vs. $B$ at temperatures $T_{g}$
in the structures "n1" ($I$-2-1, $V$-3-4, Fig. \ref{f8}), "sl5"
($I$-3-4, $V$-2-1, Fig. \ref{f9}), "sl6" (circuit a: $I$-3-4,
$V$-2-1, Fig. \ref{f10}(a)), "sl6" (circuit b: $I$-2-1, $V$-3-4,
Fig. \ref{f10}(b)).

\begin{table*}
\caption{\label{t7} Expected and fitted values used to fit the
experimental switching current vs. $B$ at temperatures $T_{g}$ to
the function $I_{c1}(T_{g},B)$ (\ref{e14}) in low fields in the
structures "n1" (Fig. \ref{f8}), "sl5" (Fig. \ref{f9}) and to the
functions $I_{ck}(T_{g},B)$ (\ref{e14}), where $k=1, 2$, in low
and intermediate fields in the structures "sl6" (Fig.
\ref{f10}(a)-(b)). $I_{cne}$ denotes the expected switching
current calculated for the individual narrow wire of the structure
at temperatures $T_{g}$ in $B=0$\,G. The meanings of other
notations are explained in the title of Table \ref{t3}.}
\centering %
\begin{tabular}{cccccccccc}
\hline
data & $T_{g}$, & $\xi(T_{g})_{e}$, & $\xi_{f1}$, & $\xi_{f2}$, & $w_{1}$, & $ w_{2}$, & $I_{cn}$, & $I_{cf1}$, & $I_{cf2}$,  \\
 & K & $\mu$m & $\mu$m &  $\mu$m &  $\mu$m &  $\mu$m &  $\mu$A & $\mu$A &  $\mu$A  \\
\hline
n1 & 1.343 & 0.64 & 0.60 & & 0.46 &  & 9.0 & 9.0 & \\
sl5 &1.289 & 0.58 & 0.48 &  & 0.40 &  & 5.5 & 9.2 & \\
sl6(a) &1.403 & 0.69 & 0.69 & 0.44 & 0.20 & 0.39 & 2.1 & 2.1 & 3.7 \\
sl6(b) &1.399 & 0.66 & 0.69 & 0.43 & 0.20 & 0.60 & 2.4 & 2.4 & 3.8 \\
\hline
\end{tabular}
\end{table*}

Fig. \ref{f8} shows the measured switching (squares) and retraping
(circles) currents vs. $B$ at $T_{g}=1.343$ K in the "n1"
structure ($I$-2-1, $V$-3-4). The experimental switching current
is fitted to the function $I_{c1}(T_{g},B)$ (\ref{e14}) in low
fields $B=0-30.61$ G (solid line 1a). The retrapping current is
significantly smaller than the switching current and weakly
depends on the field at $B=0-31$ G. In the fields $B=31.73-45.57$
G, the retrapping current coincides with the switching current
(solid line 1b). The dash-dotted line 1e is the expected switching
current vs. $B$ at $T_{g}=1.343$ K, calculated with the help of
the function $I_{GL}(T,B)$ (\ref{e6}), into which the
corresponding parameters from Table \ref{t7} are placed.

In low fields, lines 1a and 1e are located close to each other. In
high fields, a sharp drop in the switching current is observed,
followed by a slight decrease in the current as the field
increases (solid line 1b). The solid line 1b shows the area of
equilibrium fluctuations of the superconducting order parameter.

The fitted value of $\xi_{f1}$ for the "n1" structure is close to
the expected coherence length $\xi(T_{g})_{e}$ (Table \ref{t7}).
When calculating $\xi(T_{g})_{e}$, $T_{c}(0.5)=1.407$ K is taken
(Table \ref{t2}).

To calculate the expected switching current
$I_{cne}=I_{cne}(T_{g},B=0)$ at temperature $T_{g}$ in $B=0$\,G of
the "n1" structure (Table \ref{t7}),  the function $I_{GL}(T)$ is
used (\ref{e2}), where the fitted parameters are taken from Table
\ref{t2}. The measured switching current
$I_{cnm}=I_{cnm}(T_{g},B=0)$ is close to $I_{cne}$ and practically
coincides with the fitted switching current $I_{cf1}$.

Fig. \ref{f9} shows the switching (squares) and retrapping
(circles) currents as functions of the magnetic field measured in
the "sl5" structure ($I$-3-4, $V$-2-1) at $T_{g}=1.289$ K, and the
solid line 1a is the fit of the experimental switching current to
the function $I_{ck}(T_{g},B)$ (\ref{e14}), where $k=1$ in low
fields. The fitted values of $I_{cfk}$, $T_{g}$, $\xi_{fk}$,
$w_{k}$ needed to plot the function are given in Table \ref{t7}.
In low fields, the retrapping current is significantly smaller
than the switching current and depends weakly on the field. In
high fields, the retrapping current coincides with the switching
current (solid line 1b).

We plotted the expected switching current vs. $B$ at $T_{g}=1.289$
K in the two-width "sl5" ($I$-3-4, $V$-2-1) structure by
neglecting the influence of the wide current wires of the
structure on the electron transport and using the equation for the
GL critical current in a single narrow wire of width $w_{n}=0.23$
$\mu$m $I_{GL}(T, B)$ (\ref{e7}), into which the corresponding
parameters from Tables \ref{t1}, and \ref{t7} are inserted
(dash-dotted line 1e, Fig. \ref{f9}). The switching current vs.
$B$ measured at $T_{g}=1.289$ K (squares, Fig. \ref{f9}) differs
radically from the expected switching current. Line 1e intersects
the field coordinate axis at $B=87.6$ G. However, the experimental
switching current is kept up to the field of $B=132.1$ G. The
non-zero critical current in high fields (solid line 1b) is due to
the proximity effect near the junction line of the wide and narrow
wires and equilibrium fluctuations of the superconducting order
parameter.

Fig. \ref{f9} shows that the steepness of the solid line 1b is
less than the steepness of the solid line 1a. The field $B=47.3$
G, corresponding to the intersection of lines 1a and 1b, is
slightly less than the critical field $B=63.9$ G, at which the
individual (without taking into account the influence of the
proximity effect) wide wire of the "sl5" structure should
transition to the normal state. This field is defined by the
requirement $T_{c}(B)=T_{g}=1.289$ K, where $T_{c}(B)$ is given by
equation (\ref{e10}), into which the corresponding values of
$w_{1}= w_{b}=w_{w}=0.40$ $\mu$m, $\xi(0)=0.128$ $\mu$m,
$T_{c}=T_{r}^{cf}=1.405$ K are substituted (Tables \ref{t1}, and
\ref{t2}). The fitted width $w_{1}=0.40$ $\mu$m used in the
fitting function $I_{c1}(T_{g},B)$ (line 1a) indicates the
predominant influence of the electron transport of the wide
current wire of the structure on the switching current vs. $B$ in
low fields $B=0-47.3$ G (Table \ref{t7}).

Table \ref{t7} shows that the relation
$\xi(T_{g})_{e}/\xi_{f1}=1.20$ is fulfilled. When calculating
$\xi(T_{g})_{e}$, we took $T_{c}(0.5)=1.355$ K (Table \ref{t2})
for the "sl5" structure (Fig. \ref{f9}). The expected switching
current $I_{cne}=I_{cne}(T_{g},B=0)$ of the "sl5" structure (Fig.
\ref{f9}) was determined by the equation
$I_{GL}(T)=I_{GL}(0)(1-T/T_{c})^{3/2}$, where $I_{GL}(0)=506$
$\mu$A is the theoretical GL critical current at $T=0$ K for the
individual narrow wire of the structure with the width of 0.23
$\mu$m , $T_{c}=T_{c}(0.5)=1.355$ K. However, the fitted GL
critical current $I_{GL}^{f}(0)=860$ $\mu$A at $T=0$ K almost
coincides with the theoretical GL critical current $I_{GL}(0)=893$
$\mu$A (Table \ref{t2}), calculated for the individual wide wire
of the "sl5" structure with a width of 0.40 $\mu$m. $I_{cnm}$ is
almost twice as large as $I_{cne}$ and coincides with $I_{cf1}$.

Fig. \ref{f10}(a) shows the switching (squares) and retrapping
(crosses) currents as functions of magnetic field measured in the
"sl6" structure (a) ($I$-3-4, $V$-2-1) at $T_{g}=1.403$ K, and the
solid lines 1a, 1b are the fits of the experimental switching
current vs. $B$ to the corresponding functions $I_{ck}(T_{g},B)$
(\ref{e14}), where $k=1, 2$ in low, intermediate fields. The solid
line 1c (drawn for the eye) indicates the measured switching
current vs. $B$ (squares), in high fields. In low and intermediate
fields, the retrapping current is significantly less than the
switching current and is almost independent of the field. In high
fields, the retrapping current coincides with the switching
current (solid line 1c). The solid line 1c shows the area where
the wide wire is in its normal state.

Fig. \ref{f10}(a) shows that the steepness of the solid line 1b is
greater than the steepness of the solid line 1a, and the steepness
of line 1c is less than the steepness of line 1a. In this case,
the inequality $I_{cf2}>I_{cf1}$ is satisfied (Table \ref{t7}).
The field $B=53.0$ G, corresponding to the intersection of lines
1b and 1c, practically coincides with the critical field $B=52.9$
G, at which the individual (without taking into account the
influence of the proximity effect) wide current wire of the
structure should transition to the normal state. This field is
determined by the requirement $T_{c}(B)=T_{g}=1.403$ K, where
$T_{c}(B)$ is given by equation (\ref{e10}), in which the
corresponding values $w_{2}=w_{b1}=w_{w}=0.39$ $\mu$m,
$\xi(0)=0.120$ $\mu$m, $T_{c}=T_{r}^{cf}=1.472$ K are placed
(Tables \ref{t1}, and \ref{t2}). In Table \ref{t7}, the fitted
widths $w_{1}= w_{a}=w_{n}=0.20$, $w_{2}= w_{b1}=w_{w}=0.39$
$\mu$m used in the fitting functions $I_{c1}(T_{g},B)$ (line 1a),
$I_{c2}(T_{g},B)$ (line 1b) (\ref{e14}), respectively, indicate
the long-range dominant influence of the physical parameters of
the wide current wire of the structure on the switching current
vs. $B$ in the intermediate field range of $B=48.5-53.0$ G. While
in low fields, the switching current vs. $B$ is determined mainly
by the influence of the physical parameters of the narrow current
wire.

We found that the wide current wire ($w_{b1}=0.39$ $\mu$m),
located at a significant distance from the central section of the
"sl6" structure (Fig. \ref{f10}(a), $I$-3-4, $V$-2-1), affects the
electron transport in the central section of the structure.

Fig. \ref{f10}(b) shows the switching (circles) and retrapping
(asterisks) currents as functions of magnetic field measured in
the "sl6" structure (b) ($I$-2-1, $V$-3-4) at $T_{g}=1.399$ K, and
the solid lines 2a, 2b are the fits of  the experimental switching
current to the corresponding functions $I_{c1}(T_{g},B)$
(\ref{e14}) in low fields $B=0-25.1$ G, $I_{c2}(T_{g},B)$
(\ref{e14}) in the intermediate field range of $B=25.1-36.8$ G.
The solid line 2c (drawn for the eye) indicates the measured
switching currents vs. $B$ (circles), in high fields. The
retrapping current is significantly less than the switching
current and is practically independent of the field at $B=0-65$ G.
In fields $B=40-133$ G, the retrapping current coincides with the
switching current (solid line 2c).

We calculated the expected switching current vs. $B$ at
$T_{g}=1.399$ K in the three-width "sl6" (b) ($I$-2-1, $V$-3-4)
structure, neglecting the influence of wide current and potential
wires of the structure on electron transport and using the
equation for the GL critical current in a single narrow wire of
width $w_{n}=0.20$ $\mu$m $I_{GL}(T, B)$ (\ref{e6}), into which
the corresponding physical parameters from Tables \ref{t1}, and
\ref{t7} are inserted (dash-dotted line 2e, Fig. \ref{f10}(b)). A
large difference is observed between the experimental (circles,
Fig. \ref{f10}(b)) and expected switching currents vs. $B$ at
$T_{g}=1.399$ K. Line 2e practically coincides with the function
$I_{c1}(T_{g},B)$ (\ref{e14}) in low fields $B=0-25.1$ G and
intersects the field axis at $B=90$ G. However, the experimental
switching current in high fields (the solid lines 1c and 2c) is
kept up to the $B=140$ G field. The non-zero critical current is
due to the proximity effect near the junction line of the wide and
narrow wires and equilibrium fluctuations of the superconducting
order parameter in the narrow wire.

Fig. \ref{f10}(b) shows that the steepness of the solid line 1b is
greater than the steepness of the solid line 1a, and the steepness
of line 1c is less than the steepness of line 1a. In this case,
the inequality $I_{cf2}>I_{cf1}$ is satisfied (Table \ref{t7}).
The field $B=36.8$ G, corresponding to the end of line 1b,
practically coincides with the critical field $B=35.24$ G, at
which the individual (without taking into account the influence of
the proximity effect) wide current wire of the structure should
transition to the normal state. This field is determined by the
requirement $T_{c}(B)=T_{g}=1.399$ K, where $T_{c}(B)$ is given by
equation (\ref{e10}), in which the corresponding values
$w_{b2}=w_{w}=0.60$ $\mu$m, $\xi(0)=0.120$ $\mu$m,
$T_{c}=T_{r}^{cf}=1.472$ K are placed (Tables \ref{t1}, and
\ref{t2}). In Table \ref{t7}, the fitted widths
$w_{1}=w_{a}=w_{n}=0.20$, $w_{2}= w_{b2}=w_{w}=0.60$ $\mu$m used
in the fitting functions $I_{c1}(T_{g},B)$ (line 1a),
$I_{c2}(T_{g},B)$ (line 1b) (\ref{e14}), respectively, indicate
the dominant influence of the wide current wire of the structure
on the switching current vs. $B$ in the intermediate field range
of $B=25.1-36.8$ G. While the influence of the narrow current wire
determines the switching current vs. $B$ in the low $B=0-25.1$ G
fields.

We found that the wide current wire ($w_{b2}=0.60$ $\mu$m),
located at a great distance from the central section of the "sl6"
structure (Fig. \ref{f10}(b), $I$-2-1, $V$-3-4), affects the
electron transport in the central section of the structure.

Table \ref{t7} shows that for the "sl6" structure (Fig.
\ref{f10}(a)-(b)) the requirements $\xi(T_{g})_{e}/\xi_{f1}\approx
1$ and $\xi_{f2}<\xi_{f1}$ are satisfied. When calculating
$\xi(T_{g})_{e}$, we took the fitted critical temperature
$T_{GL}^{cf}=1.447$ K instead of $T_{c}(0.5)=1.463$ K (Table
\ref{t2}).

To calculate the expected switching currents
$I_{cne}=I_{cne}(T_{g},B=0)$ of the "sl6" structure (Fig.
\ref{f10}(a)-(b)) at temperatures $T_{g}$ in $B=0$\,G, we used the
functions $I_{GL}(T)$ (\ref{e2}), where the fitted parameters are
given in Table \ref{t2}. The measured switching current
$I_{cnm}=I_{cnm}(T_{g},B=0)$ at temperatures $T_{g}$ in $B=0$\,G
is close to $I_{cne}$ and practically coincides with $I_{cf1}$.

\section{Conclusions}

We measured the critical switching and retrapping currents as
functions of temperature $T$, close to $T_{c}$, in the zero field
in thin-film long constant-, two- and three-width
quasi-one-dimensional superconducting aluminum structures of
various geometries. The critical currents were determined from the
appearance and disappearance of dc voltage in the central short
section of the structure with increasing and decreasing direct
current. In the zero field, a distinctive feature of the studied
structures is that the critical temperature of the individual wide
wire of the structure is higher than the critical temperature of
the individual narrow wire of the structure.

We found that the temperature-dependent switching current of long
two- and three-width quasi-one-dimensional superconducting
aluminum structures in the zero magnetic field is nonlocal. The
switching current depends on electron transport in the area that
contains the junction line between the narrow and wide wires. The
experimental zero-field switching current as a function of
temperature is fitted to two functions. At lower temperatures, the
fitting function is the temperature-dependent Ginzburg-Landau
critical current. At higher temperatures, the fitting function is
the linear temperature dependence of the Josephson critical
current. In the uniform constant-width structure, the experimental
switching current vs. $T$ is fitted within the framework of the
Ginzburg-Landau theory to a single function at practically all
temperatures.

In a structure with a narrowing length equal to the
temperature-dependent superconducting coherence length $\xi(T)$,
we unexpectedly found that the experimental switching current is
determined by the wide wire of the structure. Moreover, the
switching current density in the narrowing is almost twice the GL
critical current density.

In addition, we measured the switching and retrapping currents as
functions of the magnetic field $B$ normal to the substrate
surface at a given temperature $T_{g}$ close to $T_{c}$ in the
five structures. In fields from zero to the maximum critical field
at which the wide wire of the structure should return to its
normal state, the experimental switching current vs. $B$ is fitted
to two functions within the framework of the Ginzburg-Landau
theory.

In intermediate fields, a situation arises when
magnetic-field-dependent critical temperature $T_{c}(B)$
(\ref{e10}) of the individual wide wire of the structure is lower
than the critical temperature $T_{c}(B)$ (\ref{e10}) of the
individual narrow wire.

We found that the measured switching current as a function of the
field at a given temperature is nonlocal. The switching current is
determined by the long-range influence of electron transport in
the area including the junction line of the narrow and wide
current wires.

When fitting, the two-width structure (when the applied current
flows through wide and narrow wires) was considered as single
narrow or wide wires in low or intermediate fields, respectively.

We found that for two- and three-width structures, the
experimental switching current vs. $B$ differs from the calculated
expected switching current vs. $B$. For the constant-width
structure, the experimental and expected currents as functions of
$B$ at a given temperature are practically identical in almost all
fields.

At low fields (when current flows through the narrow and wide
wires of the structure), the switching current is determined
primarily by electron transport in the narrow current wire.

In the intermediate field range, a sharp decrease in the switching
current with increasing field occurs. This is due to the
predominant nonlocal influence of the physical parameters of the
wide wire on the magnetic-field-dependent switching current in
two-width structure.

In high fields, the switching current coincides with the
retrapping current. The switching current is small, decreases
slightly with increasing field, and exists in fields higher than
the maximum critical magnetic field in a quasi-one-dimensional
superconducting wire. In high fields, the non-zero critical
current is due to the proximity effect near the junction line of
the wide and narrow wires and equilibrium fluctuations of the
superconducting order parameter.

Our work is experimental. We do not claim to provide a rigorous
theoretical explanation of the experimental results. A modified
Ginzburg-Landau theory accounting for the inhomogeneous
distribution of the order parameter, or the use of the Usadel
formalism  \cite{usadel}, might help provide a rigorous
explanation of the results.

For quasi-one-dimensional superconducting aluminum wires of the
same thickness in a zero magnetic field, it was found that the
smaller the wire width, the lower the critical switching current
density at a given temperature \cite{kuznphysicaccrit22}. This
finding has already provided insight into the operation of a
superconducting ac voltage rectifier in low fields
\cite{dubjetplet03}, \cite{kuznphysicacshift22}. We believe that
the results of our study, together with the results of
\cite{kuznphysicaccrit22}, will help elucidate the mechanisms
behind insufficiently understood effects, such as the "resistive
transition anomaly" \cite{art16}, \cite{art17}, \cite{art18},
\cite{art19}, \cite{art20}, \cite{art21}, \cite{art22},
\cite{art23}, negative magnetoresistance in low magnetic fields
\cite{art24}, \cite{art25}, \cite{art26}, and the "antiproximity"
effect \cite{art27}, \cite{art28}, \cite{art29}. Furthermore,
these results will aid in understanding the operation of a
superconducting ac voltage rectifier in high fields
\cite{dubjetplet03}. We believe that our results will help in the
creation of an asymmetric micro-SQUID \cite{art7}.

\section{Acknowledgments}

We thank A. Firsov for fabricating the aluminum structures. This
work was completed and financially supported under STATE TASK No.
075-00296-26-00.



\end{document}